\documentclass[prd,amsmath,amsfonts,preprint,12pt,superscriptaddress]{revtex4}

\usepackage{epsfig}
\newcommand{\coH}[2]{\mathcal H^{#1} \left( #2 \right) }
\newcommand{\mc}[1]{\mathcal{ #1 }}
\newcommand{\er}[1]{(\ref{#1})}
\newcommand{\diff}{ \mbox{d}} 

\newcommand{\nonC}{\mathbf{R}^{1,3}}
\newcommand{\isC}{\mc M}

\newcommand{\difs}{ \mbox{d}^\dagger}
\newcommand{\fullspace}{ \nonC \times \isC }
\newcommand{\real}[1]{ \mathbf{R}^{#1}}
\newcommand{\torus}[1]{ \mathbf{T}^{#1}}
\newcommand{\sphere}[1]{ \mathbf{S}^{#1}}
\newcommand{\Zto}[1]{ \mathbb{Z}^{#1} }
\newcommand{\Zmod}[1]{ \mathbb{Z}_{#1} }
\newcommand{\CP}[1]{ \mathbb{C} \mbox{P}^{#1} }
\newcommand{\smodz}{\sphere{1}/\Zmod{2}}

\begin{document}

\title{Controlling Chaos through Compactification in\\ Cosmological Models with 
a
Collapsing Phase}
\author{Daniel H. Wesley}
\email{dhwesley@princeton.edu}
\affiliation{Joseph Henry Laboratories, Princeton University, Princeton NJ, 
08544}
\author{Paul J. Steinhardt}
\affiliation{Joseph Henry Laboratories, Princeton University, Princeton NJ, 
08544}
\author{Neil Turok}
\affiliation{DAMTP, CMS, Wilberforce Road, Cambridge, CB3\,0WA, UK}

\date{29 July 2005}

\begin{abstract}
\noindent We consider the effect of compactification of extra dimensions on the
onset of classical chaotic ``Mixmaster" behavior during 
cosmic contraction.  
Assuming a universe that is well--approximated as a four--dimensional
 Friedmann--Robertson--Walker model (with negligible Kaluza--Klein excitations)
 when the contraction phase begins, 
we identify compactifications that allow a smooth contraction and delay the 
onset of chaos until arbitrarily close to the big crunch.
These compactifications are defined by the de Rham cohomology (Betti numbers) 
and Killing vectors of the compactification manifold.
We find compactifications that control chaos 
in vacuum Einstein gravity, as well as in string theories with $\mc N = 1$ 
supersymmetry and M--theory.
In models where chaos is controlled in this way, the universe can remain 
homogeneous
and flat until it enters the quantum gravity regime.  At this point, the 
classical
equations leading to chaotic behavior can no longer be trusted, and quantum 
effects
may allow a smooth approach to the big crunch and transition into a subsequent
expanding phase.
Our results may be useful for constructing cosmological models with contracting
phases, such as the ekpyrotic/cyclic and pre--big bang models.
\end{abstract}

\maketitle
\tableofcontents
\newpage

\section{Introduction}

The behavior of spacetime near a big crunch singularity has been a topic of
research for many decades.
The classic studies of Belinskii, Khalatnikov, and Lifshitz
(BKL) \cite{Lif63,Bel70,Bel73} and others 
\cite{Dem85,Dam00,Dam00A,Dam02,Dam02A,Dam02B,Mis69,MTW,Cor97} 
have shown that the contraction to the crunch can proceed either
smoothly or chaotically.  Chaos arises when the universe is unstable to 
 small inhomogeneities and anisotropies in curvature or matter fields.  
These ``dangerous"
perturbations grow and eventually dominate the dynamics, driving the universe
to an anisotropic state, expanding along some axes and contracting along others.  
The axes and their rates of contraction jump to new values 
when new curvature or matter terms grow to dominate.  The jumps generally 
repeat an infinite number
of times before the big crunch itself.  The chaotic, oscillatory evolution to 
the
big crunch is often known as ``Mixmaster" behavior 
or ``BKL oscillations."

Models that generalize four dimensional Einstein gravity have been
classified by whether they necessarily exhibit chaotic behavior near
a big crunch.
This classification is established assuming ``generic" initial
conditions, in which there is a finite, but possibly small, energy density in 
all fields 
present in the model.   
Additionally, one assumes that the classical Einstein equations remain
valid up to the big crunch itself.
Under these assumptions, it has been shown that vacuum Einstein gravity in 
spacetime
dimension less than eleven, as well as all 
uncompactified ten dimensional string theories and
M--theory, will inevitably suffer chaos as the big crunch is approached
\cite{Dem85,Dam00,Gib05}.
So too will
Einstein gravity containing only perfect fluids with equation of state $w < 1$, 
where
$w = p/\rho$ is the ratio of the fluid's pressure to its energy density.
On the other hand, there are cases where chaos is not inevitable.  
Among these are Einstein gravity with a free massless scalar
field, leading to $w=1$, 
and a universe containing a perfect fluid with equation of state $w \ge 1$ 
\cite{Eri03}.

The presence of chaos during gravitational collapse is a potential problem for
 cosmological models with a big crunch/big bang
transition, such as the ekpyrotic/cyclic and pre big--bang scenarios
\cite{Kho01A,Ste02,Ven00,Gas02}.  
In these models, the universe undergoes collapse to a big crunch, followed
by a transition to the conventional big bang and subsequent expansion.
It is assumed that, during the collapsing phase, the universe is 
nearly homogeneous and isotropic, with a scale invariant perturbation spectrum.
  BKL oscillations arising during the collapsing
phase would destroy homogeneity and isotropy, producing a chaotic spacetime
with structure down to arbitrarily small scales.  In this situation, a 
big crunch/big bang transition is unlikely to be describable in a deterministic
manner, and it is questionable whether a homogeneous and isotropic universe
with the long range correlations required by observations could emerge 
\cite{Pee72}.
Thus, avoiding chaos is an essential feature of cosmological models with a 
collapsing phase.

We take the point of view that avoiding chaos all the way to the the big crunch
is too restrictive a condition for viable cosmological models.  
One expects classical general relativity to break down at a small but finite 
time before the big crunch is reached, perhaps of order a Planck time $t_{PL}$
or string time $t_S$.  
After this point, quantum effects become significant and we can no longer trust
the classical physics that predicts a chaotic approach to the big crunch.  
Provided the universe evolves smoothly and
non--chaotically until $t_{PL}$, it is conceivable that quantum gravity effects
allow the universe to pass smoothly through the big crunch and into a subsequent 
expanding phase.  For example, recent work \cite{TPS04} has revealed that the 
degrees of freedom present in string and M--theory (extendend objects such as
 strings and
branes) can evolve smoothly through certain types of big crunch singularities.
This suggests that chaotic behavior is absent in the quantum gravity regime, and 
furthermore a nonsingular transition from a big crunch to a big bang is 
possible.

Our focus in the present work is on the classical evolution of the universe 
before $t_{PL}$, and whether it is possible for it to evolve smoothly so long 
as we can trust the classical equations of motion.
The evolution of the universe through the subsequent quantum regime,
and the transition to an expanding phase, are important though 
unsettled issues.  In this work we have nothing to add on these topics.
However, if chaos can be controlled  during the classical evolution of the
universe, there remains the hope that the subsequent quantum evolution will
preserve the long range correlations, isotropy and homogeneity so essential for 
cosmology.

In this paper, we consider models that are well described by a classical, 
four--dimensional effective field theory long before the big crunch.
The four--dimensional metric is that of a nearly 
homogeneous and isotropic Friedmann--Robertson--Walker (FRW) universe, 
with small perturbations to the metric, matter and Kaluza--Klein fields, and
with the Hubble radius $H^{-1}$  much larger than the compactification length 
scale $R_c$.  
We  study the evolution of chaotic behavior as the universe contracts, 
including the effects of all massive Kaluza--Klein modes, and thus all of the 
degrees of freedom of the higher--dimensional theory.
We show that the emergence of chaos can be controlled provided the 
compactification
manifold $\isC$ satisfies certain topological conditions.  For these topologies,
the dangerous perturbations that formerly led to chaos acquire masses of order 
the inverse of the compactification length scale, $m \sim 1/R_c$.
The presence of mass terms slows the growth of energy density 
in these fields, and prevents them from becoming cosmologically relevant so long
 as the Hubble parameter is larger than their mass.
When the time until the big crunch becomes less than $R_c$, or equivalently $m < 
H$, the suppression ceases to operate, and the energy density in the dangerous 
modes can grow at their usual unsuppressed rate.
However, since the energy density in
these heavy modes has been greatly suppressed relative to light modes up to this
point,
they cannot dominate the 
energy density until the universe has contracted further by an
exponential factor.
Typically,  the
massive 
modes do not dominate before we enter the quantum gravity regime 
at roughly $t_{PL}$, at which point the classical  evolution equations cannot be 
trusted. 
In these circumstances,
we say that chaos has been ``controlled."

In this paper, we focus on a classical effect which reduces the 
importance of chaos in compactified models.  This is especially relevant
for models, such as string-- or M--theory, in which the compactification of
extra dimensions is an essential element.  An excellent example is given by
eleven--dimensional supergravity, whose bosonic sector contains a four--form
field strength in addition to the metric.  Without the four--form, pure
eleven--dimensional gravity is not chaotic.  For some choices of the topology
of the compactification manifold, it is
possible to remove the light modes of the four--form field.
For these topologies, the
previously chaotic eleven--dimensional supergravity theory will behave 
like the non--chaotic eleven--dimensional pure gravity theory.

More generally, when 
one studies a fully uncompactified model, one finds that
chaos arises from dangerous modes that are nearly spatially homogeneous
along all dimensions.  The energy density in these modes scales rapidly
enough to dominate the universe and cause chaos.  
As we detail below, for some choices of the compactification manifold, the
classical equations of motion forbid spatially homogeneous modes along the
compact directions, for topological reasons.
In the four--dimensional effective theory, this is reflected in the appearance
of large mass terms for the associated degrees of freedom.
As we will show, as long as the four--dimensional Hubble parameter is larger
than their mass, the energy density in these massive modes grows far more 
slowly than the energy density in the light modes which dominate the dynamics.
As the Hubble radius falls below the compactification length scale, the energy
in the massive modes begins to grow more quickly, but due to its relative
suppression, it remains dynamically irrelevant all the way to the Planck 
time.
In the present work, we focus exclusively on the classical evolution of fields
and find that it suffices to control chaos.
It is possible that quantization, by imposing a further constraint on the
initial perturbations, would further suppress chaos.  The quantum production
of heavy Kaluza--Klein modes is, na\"{i}vely at least, completely negligible
all the way to the Planck time.

In this work, we consider both pure Einstein gravity and 
models with additional matter fields. We focus on $p$--form matter fields, with
exponential couplings to a scalar ``dilaton" field $\phi$, defined by the 
action,
\begin{equation}\label{eq:pFormDef}
S = -\frac{1}{(p+1)!} \int e^{\lambda\phi} F_{p+1}^2 \, \sqrt{-G} \; d^D x,
\qquad F_{p+1} = \diff A_p,
\end{equation}
with $\lambda$ a constant \cite{conventions}.  Supergravity and string models
commonly include $p$--form fields with couplings of this type.
In the following, we will always use ``$p$'' to denote the number of indices on 
the 
gauge potential $A_p$.
While many models that generalize four dimensional Einstein gravity contain
fermionic fields, throughout this work we will focus exclusively on the bosonic
sector.  We will also neglect more exotic terms in the $p$--form action such as 
Chapline--Manton
couplings and Chern--Simons terms, which at any rate we do not expect to be 
relevant
for chaos \cite{Dam02}.
Another important type of matter, the perfect
fluid, has been discussed elsewhere \cite{Eri03}, and is not affected by the 
compactification of extra dimensions.

In Section \ref{s:review}, we review some results regarding the emergence of 
chaos
during gravitational collapse that are required in later sections.  
This section is primarily concerned with distinguishing 
between chaotic and non--chaotic models.
Most importantly, we define the gravitational, electric and magnetic stability
conditions that must be satisfied if a theory is to avoid chaotic behavior. In
Section \ref{s:cc}, we introduce some key aspects of  models with 
``controlled chaos," that are the subject of the current work.
We establish that giving masses to dangerous modes prevents them from causing
chaos.  We then describe how the masses of these fields are determined by the 
compactification
manifold.
With these results established, we present our central new results in 
Section \ref{s:msc}.  These are the ``selection rules" that determine a
subset of stability conditions that must be satisfied in order to control chaos.
These rules express the precise correspondence between the properties of the 
compactification manifold, and the chaotic behavior of the 
lower--dimensional theory after compactification.
In Section \ref{s:examples}, we give examples with specific 
compactification manifolds that are able to control chaos in vacuum Einstein 
gravity, 
string theories with $\mc N = 1$ supersymmetry, and M--theory.
We summarize our conclusions in Section \ref{s:conclusions}, and suggest some 
areas for further research.

\section{Review of Gravitational and $p$--form Chaos}\label{s:review}

The essential principle underlying the emergence of chaos is 
that,  near a big crunch, 
solutions to the Einstein equations are strongly unstable to perturbations.  
This
phenomenon may be studied using a suitably general metric, such as the 
generalized
Kasner metric,
\begin{equation}\label{eq:genKasner}
ds^2 = -dt^2 + \sum_{j=1}^{D-1} t^{2p_j}  (\sigma_j)^2,
\end{equation}
where $D$ is the dimension of spacetime, the $\sigma_j = \sigma_{j \mu}(x) \diff 
x^\mu$ 
are independent of time, and the big crunch occurs at $t=0$.  
The Kasner exponents $p_j$ may be spatially varying, but upon substituting 
\er{eq:genKasner} into the Einstein equations, one finds the $p_j$ are
constrained by the Kasner conditions,
\begin{equation}\label{eq:kasnerCond}
\sum_{j=1}^{D-1} p_j = 1, \qquad \sum_{j=1}^{D-1} p_j^2 = 1.
\end{equation}
The first condition defines the \emph{Kasner plane}, the second defines the 
\emph{Kasner sphere}, and we may term their intersection the \emph{Kasner 
circle}.
For these anisotropic metrics it will be convenient to define the 
analogue of the Hubble parameter in the conventional, isotropic
FRW solution.  Using \er{eq:genKasner}, the metric on equal time
hypersurfaces is given by 
$h_{\mu\nu} = \sum_j t^{2p_j} \sigma_{j\mu} \sigma_{j\nu}$,
allowing us to define,
\begin{equation}
H = \frac{d}{d t} \log \sqrt{h}, \qquad \mbox{where} \qquad h = \det h_{\mu\nu}.
\end{equation}
The first Kasner condition  implies that $H = t^{-1}$ for any choice of the 
Kasner exponents.  We will find this Hubble parameter a useful guide to the 
typical 
dynamical timescale of the gravitational field.

The Kasner metric \er{eq:genKasner} has been widely used as a tool to understand
the behavior of ``generic" spacetimes near a big crunch singularity 
\cite{Lif63,Bel70,Bel73,Dem85,Dam00,Dam00A,Dam02,Dam02A,Dam02B,Cor97}.  It is an
approximation, to leading order in $t$, 
of an exact solution of the Einstein
equations, in which we have neglected the influence of spatial derivatives and 
curvature terms.  To check whether this approximation is
consistent, we  substitute the Kasner metric into the Einstein equations, and 
check
that terms corresponding to spatial derivatives and curvature appear at 
subleading order in $t$.  This corresponds to spatial derivatives and curvature
terms becoming irrelevant as the big crunch is approached.
It has been shown rigorously \cite{Dam02} and numerically \cite{Cor97} that,
when these terms are subleading, solutions to Einstein's equations
asymptotically approach Kasner form as $t \to 0$.

A useful feature of generalized Kasner universes is that
the conditions determining whether the curvature terms are irrelevant can be
expressed entirely in terms of the Kasner exponents.
The Einstein tensor for the Kasner metric \er{eq:genKasner} may be split into 
purely 
time derivative terms, and terms arising from the curvature of the
spatial slices.  Details of this decomposition are given in \cite{Lif63,Bel70}.  
One finds that the time derivative terms
all scale as $t^{-2}$, while the terms arising from spatial curvature
scale as $t^{-2 ( p_i + p_j - p_k )}$, for all  triplets $i,j,k$.
Dangerous components of the curvature are those whose corresponding terms in the
Einstein equations grow more rapidly than $t^{-2}$, as $t\to 0$,
thus invalidating the Kasner
approximation.  These dangerous curvature terms are absent provided that the 
\emph{gravitational stability conditions},
\begin{equation}\label{eq:gravStability}
p_i + p_j - p_k < 1, \qquad \mbox{all triples} \quad i,j,k ,
\end{equation}
are satisfied.  If the stability conditions are satisfied, then the evolution is 
guaranteed to be smooth and Kasner--like all the way to the big crunch.
These conditions turn out to be very restrictive; in the absence of matter, it 
is only
 possible to simultaneously satisfy 
(\ref{eq:kasnerCond}) and (\ref{eq:gravStability}) when the spacetime 
dimension is greater than ten \cite{Dem85}.

When the gravitational stability conditions are not satisfied, then
spacetime will exhibit chaotic behavior.  Violation of the
gravitational stability conditions \er{eq:gravStability}
means that the generalized Kasner solution (\ref{eq:genKasner}) 
is invalid, and so we must find a different description.
One useful picture recasts the evolution of the metric in terms of 
geodesic motion of a point mass, undergoing reflections from 
a set of sharp walls \cite{Dam02A}.  
The free flight of the point between wall collisions is described by the 
Kasner metric, where the Kasner exponents give the momentum components of the 
moving point.
Collisions with  walls correspond to 
spatial curvature terms temporarily dominating the Einstein equations, and
result in a sudden change in the Kasner exponents.  The motion of the point in 
the
chamber defined by these walls is chaotic, and so the dynamics of spacetime is
as well.

It is useful to distinguish more precisely
between the various possibilities with respect to the stability conditions 
\er{eq:gravStability}.  A model is chaotic if 
the stability conditions cannot be satisfied for any choice of Kasner exponents
satisfying the Kasner conditions.
We will also term models chaotic when the stability conditions are only 
satisfied at
isolated points on the Kasner circle.  For example, 
it is possible to almost satisfy the stability conditions in any spacetime 
dimension
 with the so called 
``Milne" solutions, in which a single Kasner exponent is unity and the rest are
zero.  This leads to marginally dangerous curvature components that scale 
exactly as 
$t^{-2}$.
Thus, one might argue that if these curvature components are not dominant
initially, they will remain subdominant all the way to the crunch, and chaos
will 
not 
arise.  However, since this
occurs only for isolated points on the Kasner circle, any small perturbation of 
the
Kasner exponents away from the Milne solution results in a violation of the
stability conditions and the emergence of chaos.  These solutions are thus not 
practically useful
from the perspective of avoiding chaos in cosmological models, since they do 
not admit the small inhomogeneities and anisotropies that must be present in any 
physically realistic scenario.
Therefore,  
we consider a model 
non chaotic only when there exists an open region of the Kasner
circle in which all of the stability conditions are satisfied.

The presence of matter can either enhance or suppress chaos.  
Three important examples are a free massless
scalar field, $p$--form fields, and perfect fluids.  The first, a scalar field, 
suppresses chaos by modifying
 the Kasner conditions. A homogeneous, massless scalar field $\phi$ coupled to 
the Kasner metric will evolve as,
\begin{equation}
 \phi(t) = \phi_0 + p_\phi \log t,
\end{equation}
with the constants $\phi_0$ and $p_\phi$ determined by the initial conditions.
Including the stress energy from $\phi$ in the Einstein equations results in the
new Kasner conditions,
\begin{equation}\label{eq:kasnerScalarCond}
\sum_{j=1}^{D-1} p_j = 1, \qquad  \sum_{j=1}^{D-1} p^2_j = 1- p_\phi^2.
\end{equation}
While the scalar field ``Kasner exponent" $p_\phi$ enters the Kasner conditions,
it does not enter the gravitational stability conditions \er{eq:gravStability}.
It is now possible to find $p_j$ that satisfy 
these stability conditions in any spacetime dimension $D$.  For example, the
isotropic choice,
\begin{equation}
p_\phi = \sqrt{\frac{D-2}{D-1}} \qquad p_j = \frac{1}{D-1},
\end{equation}
satisfies all of the stability conditions.
Moreover, there is a finite,  open neighborhood
on the Kasner circle surrounding the isotropic solution where the stability 
conditions
are satisfied.
Essentially, there are two key properties of the scalar field that enable it to 
suppress 
chaos.
The scalar field has an isotropic stress energy tensor, and therefore
does not enhance any preexisting anisotropy
in the Kasner metric.  Also, the scalar field energy density scales as
$t^{-2}$, and thus grows sufficiently rapidly to remain relevant near the big 
crunch.

The addition of $p$--form fields coupled to $\phi$
tends to enhance chaos.  A $p$--form field strength
with $p > 1$ has an anisotropic stress energy tensor, which tends to 
enhance any preexisting anisotropy in the Kasner metric.
If the $p$--form field dominates the
energy density, then it tends to drive the universe to anisotropic oscillations 
and then 
chaos.  A homogeneous $p$--form evolves simply, and its dynamics depend on 
whether it has
a time index (an electric $p$--form) or whether all indices are spatial (a 
magnetic 
$p$--form).
Using the equation of motion and the Bianchi identity, one finds,
\begin{equation}\label{eq:homoPForm}
\left( F_{p+1} \right)^{t j_1 j_2 \dots j_{p}} \sim \frac{e^{-\lambda 
\phi}}{\sqrt{h}}, 
\qquad
\left( F_{p+1} \right)_{j_1 j_2 \dots j_{p+1}} \sim \mbox{const.}
\end{equation}
Using these solutions, we may construct the stress energy tensor for the $p$--
form 
field, and 
compare its time dependence with the $t^{-2}$ leading dependence of the 
homogeneous
terms in the Einstein equations.  Note that, unlike the scalar field, we do not
include the gravitational back--reaction from the $p$--form fields, and merely 
check 
if it remains subdominant if subdominant initially.
One finds that the $p$--form energy scales more slowly than $t^{-2}$, and 
therefore cannot cause chaos, when the following 
\emph{$p$--form stability conditions} are satisfied;
\begin{subequations}\label{eq:EBstability}
  \begin{align}
    \sum_{p}   p_j - \frac{\lambda p_\phi}{2}  &> 0 \qquad \mbox{(electric)}, \\
    \sum_{p+1} p_j - \frac{\lambda p_\phi}{2}  &< 1 \qquad \mbox{(magnetic)}.
  \end{align}
\end{subequations}
The electric conditions involve a sum of $p$ distinct Kasner exponents, 
corresponding to 
the $p$ spatial indices of an electric $p$--form field strength.  The magnetic 
conditions likewise involve a sum over $p+1$ distinct Kasner exponents.  Thus, 
for each
$p$-- or $(p+1)$--tuple of Kasner exponents, there will be a corresponding
stability condition.
The inequalities \er{eq:EBstability} are often referred to individually as
 the \emph{electric} or 
\emph{magnetic stability conditions}.  If they are violated, then 
chaos will arise.

The inclusion of a matter component with $w \ge 1$ can suppress chaos 
\cite{Eri03}.
The $w\ge 1$ fluid suppresses chaos by growing rapidly to dominate the energy 
density of the universe.  Once it dominates, the power--law Kasner solution 
\er{eq:genKasner} is replaced by a solution of the approximate form,
\begin{equation}
ds^2 = -dt^2 + t^{4/\left[(D-1)(1+w)\right]} \sum_{j=1}^{D-1} 
\exp{ \left( c_j t^{\frac{w-1}{w+1}} \right)}
(\sigma_j)^2
\end{equation}
where the $c_j$ are arbitrary constants.  When $w >1$, this solution
converges rapidly to isotropy as $t\to 0$.   Near the big crunch, this solution
may be thought of as a Kasner universe with Kasner exponents,
\begin{equation}
p_j = \frac{2}{(D-1)(1+w)}.
\end{equation}
The gravitational stability conditions are clearly satisfied in this
case, and thus chaos does not arise.  The $p$--form case is somewhat more 
subtle,
and depends on whether we realize the $w > 1$ fluid by introducing a potential 
for the
dilaton $\phi$, or as a separate matter component that does not couple to the
$p$--forms.  In either case, for any $p$--form and coupling
$\lambda$, there is a corresponding $w_{crit}(p,\lambda)$, for which chaos 
is eliminated when  $w > w_{crit}$.

\section{Massive Modes, Chaos, and Compactification}\label{s:cc}

In this section, we will distinguish a subclass of the chaotic models, those 
with
``controlled" chaos.  The salient feature of controlled chaos is that dangerous
modes are suppressed, relative to $t^{-2}$, for an adjustable epoch of cosmic
history.  In these models, dangerous modes acquire a mass $m$,
and their contribution to the total energy density of the universe is suppressed
so long as $m/H > 1$.  These masses arise through compactification, and are of 
order $R^{-1}_c$,   with $R_c$ the characteristic length scale of the
compactification manifold $\isC$.  In a universe with Kasner metric 
\er{eq:genKasner}, $H = t^{-1}$.  Thus, in these models chaos cannot emerge 
until $t \ll R_c$, 
and is is possible to ensure that chaos does not arise before the Planck time
$t_{PL}$.
In Section \ref{ss:massAndMassless}, we discuss how the growth of dangerous 
modes is suppressed by mass terms, and estimate the suppression factor.
We give a simple argument based on a scalar field in an isotropic universe,
leaving a discussion of the general $p$--form case to Appendix 1.
Sections \ref{ss:pFormSpectrum} and \ref{ss:gravitySpectrum} describe 
the mechanism by which compactification gives the required masses to $p$--form 
and 
metric degrees of freedom.  In both cases, massless modes in the lower dimension
can only exist when $\isC$ admits $p$--form or vector fields with certain 
special properties.  In Section \ref{s:msc}, we use these results to give the
``selection rules" that express the correspondence between these special 
properties
of $\isC$ and the chaotic properties of the compactified theory.

In the following, we  assume that spacetime has the form $\fullspace$, with
$\isC$ a compact manifold.  
Indices $M,N,P \dots$  denote directions in the
total spacetime $\fullspace$, while $\mu,\nu \dots$  denotes directions along
$\nonC$ and $m,n,p \dots$ along $\isC$.  $G_{MN}$  denotes the total metric, 
with
$h_{\mu\nu}$ the metric on $\nonC$ and $f_{mn}$ the metric on $\isC$.  In 
later sections we will need to distinguish between coordinate indices and
tangent space indices on $\fullspace$.  We  use $A,B,C ...$ for 
tangent space indices along the full space, $\alpha,\beta ...$ for those along
$\nonC$, and $a,b ...$ for those along $\isC$.

\subsection{Massive and massless modes}\label{ss:massAndMassless}

The suppression of massive modes in a collapsing universe may be
illustrated using the equation of state $w$.  
This is defined as the ratio of the pressure to energy density for a perfect 
fluid,
\begin{equation}
w = \frac{p}{\rho} = \frac{{T_j}^j}{-{T_0}^0},
\end{equation}
where we assume that we are in the comoving frame where the stress energy tensor 
is 
diagonal.  As we are primarily interested in
cosmological models, it is sufficient to consider the case where the universe is
an isotropic, four dimensional FRW model
 after compactification. 
Conservation of stress energy implies that the energy density $\rho$ 
of a perfect fluid with equation of state $w$ depends on the scale factor $a$ 
as,
\begin{equation}\label{eq:rhoWithA}
\rho = \rho_0 a^{-3(1+w)},
\end{equation}
where $\rho_0$ is the energy density when $a$ is unity.  
In a contracting universe,
the component with the largest $w$ grows most rapidly, and 
eventually dominates the total energy density.
A homogeneous, massive scalar field  has a perfect fluid stress
energy  with equation of state,
\begin{equation}\label{eq:w4scalar}
w = \frac{\dot \phi^2 - m^2 \phi^2}{\dot \phi^2 + m^2 \phi^2}.
\end{equation}
From  equations \er{eq:rhoWithA} and \er{eq:w4scalar} it is readily seen that 
the 
energy density of a massive scalar field
must scale more slowly than that of a massless one.  
A massless scalar field will always have $w=1$, and 
thus its energy density $\rho$ scales with $a$ as $\rho \sim a^{-6}$.
The energy density of a massive scalar will scale with an effective $w$ between 
zero 
and unity, depending on the ratio $m/H$.
When $m/H \gg 1$, far from the big crunch, the scalar field's dynamics 
is dominated by the mass term in its potential.  Using 
$\langle \cdot \rangle$ to denote the time average, the virial theorem implies 
that
$\langle \dot \phi^2 \rangle = m^2 \langle \phi^2 \rangle$, and therefore 
$w=0$ \cite{Tur83}.  
Thus the energy density in the massive field scales as $\rho \sim a^{-3}$,
far more slowly than the massless field.
Near the crunch, when $m/H \ll 1$, the mass term has a negligible effect
on the field's dynamics. In this limit,
$ \dot \phi^2  \gg m^2  \phi^2 $, and $w$ approaches
unity from below.  In this regime, the energy density in
massive and massless fields will scale identically with time.  

\begin{figure}
\epsfig{file=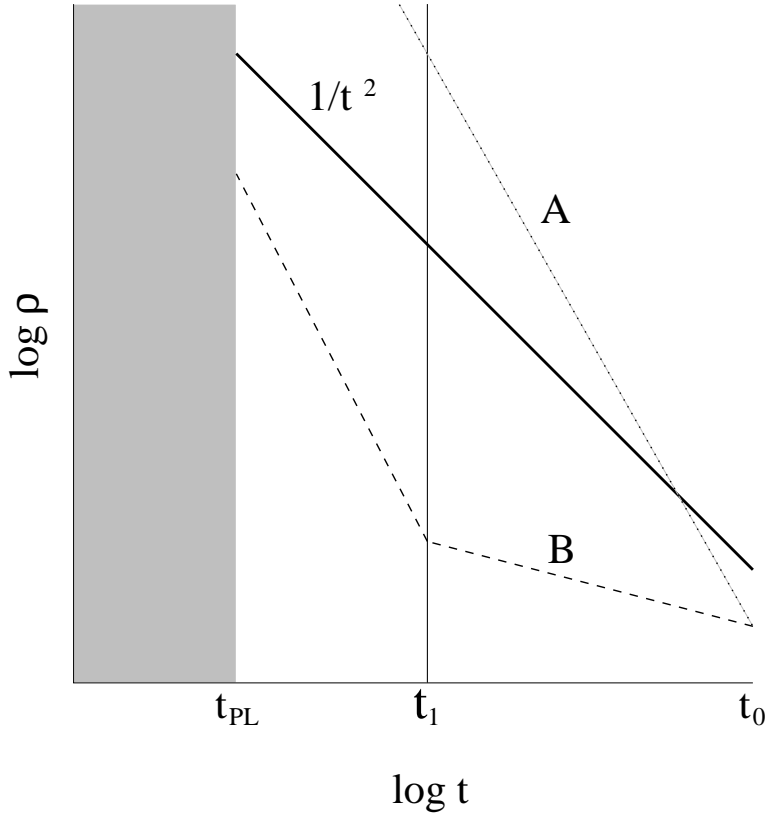,width=4in}
\caption{Scaling of the energy density in dangerous modes before and after
compactification.  Without a mass, a dangerous mode scales more rapidly than
$t^{-2}$ and eventually dominates the energy density of the universe, as 
illustrated
in {\bf A}.  With a mass, the dangerous mode scales much more slowly than
$t^{-2}$, until a time $t_1$ at which $m=H$, as seen in {\bf B}.  After this 
point,
the mode will scale as usual.  The chaos is ``controlled" when the dangerous 
energy density cannot catch up to $1/t^2$ before the Planck regime is reached at 
$t_{PL}$.}
\label{f:diffRhoScaling}
\end{figure}

Using this equation of state argument, we can estimate the exponential
suppression of the energy density in massive fields.  The basic process by
which dangerous modes are suppressed, and then grow near the big crunch is
 illustrated in Figure \ref{f:diffRhoScaling}.
Although we have discussed
only the scalar case above, the same $w=0$ scaling when $m/H > 1$ obtains for 
general $p$--form fields, as discussed in Appendix 1, and so this estimate 
applies
to those fields as well.
It is important to emphasize that, while the energy density
in massive modes is always growing, it grows more slowly than the 
$t^{-2}$ scaling required for the field to be cosmologically relevant.  
Therefore,
if we wish to estimate the importance of a given energy component,
we should consider the ratio of the energy density in the component to 
$t^{-2}$.  This quantity, $t^2\rho(t)$, may be thought of as measuring
the ratio of the
energy density in a given component to the total density, for,
\begin{equation}
t^2 \rho(t) \sim \frac{\rho(t)}{H^2} \sim \frac{\rho(t)}{\rho_{tot}(t)},
\end{equation}
where we have used Planck units.
Only when $t^2\rho(t)$ is increasing as $t \to 0$ can a given component grow to 
dominate the universe.

We begin by choosing a reference time $t_0$, at the beginning of the contracting 
phase.
We consider a model where the four--dimensional effective theory begins to
contract with a background
equation of state $\bar w$, so that,
\begin{equation}
a(t) = (t/t_0)^{\frac{2}{3(1+\bar w)}},
\end{equation}
where we have normalized $a=1$ when $t = t_0$.    
Combining this with
\er{eq:rhoWithA}, one finds that for massive modes,
\begin{equation}
t^2 \rho(t) = \left[ t^2 \rho(t) \right]_0 \cdot (t/t_0)^{\frac{2\bar w}{1+\bar 
w}},
\qquad (m/H > 1),
\end{equation}
where $\left[ t^2 \rho(t) \right]_0$ denotes 
$t^2\rho(t)$ evaluated at $t_0$.  This equation is valid up until $m/H \sim 1$, 
when the
mass terms are becoming irrelevant.  Denoting by $t_1$ the time at which
$m=H$, one finds,
\begin{equation}
\frac{t_1}{t_0} = R_c H_0,
\end{equation}
where we have taken $m = 1/R_c$.  We then have,
\begin{equation}\label{eq:rhoAtHisM}
\left[ t^2 \rho(t) \right]_1 = \left[ t^2 \rho(t) \right]_0 
\cdot (R_c H_0)^{\frac{2\bar w}{1+\bar w}}, \qquad (m/H \sim 1).
\end{equation}
This equation shows the suppression in the fractional energy density in massive
modes during the period in which $m/H > 1$.
The suppression is controlled 
by the ratio of the compactification length
scale $R_c$ to the Hubble horizon $L_H = 1/H_0$ at the beginning of the 
contracting
phase.

As we approach the big crunch, we have $m/H < 1$, and the dangerous modes can 
grow as usual.
We parameterize this growth by an exponent $\delta$, so that,
\begin{equation}\label{eq:dangerGro}
t^2 \rho(t) = (t/t_1)^{-\delta} \left[ t^2 \rho(t) \right]_1, \qquad (m/H < 1).
\end{equation}
when $\delta>0$, a mode will grow and eventually dominate the energy density of 
the universe.
One can see by comparing \er{eq:dangerGro} to our discussion in Section
\ref{s:review} that $\delta$ is merely twice the amount by which a given mode violates 
the
stability conditions.  Typically, $\delta$ will be of order one.
Now we define a time $t_{eq}$, at which the dangerous modes have grown 
sufficiently
so that the fractional energy density in dangerous modes is equal to that at the
beginning of the contracting phase, or,
\begin{equation}
\left[ t^2 \rho(t) \right]_{eq} = \left[ t^2 \rho(t) \right]_0.
\end{equation}
Using \er{eq:rhoAtHisM} and \er{eq:dangerGro}, we find,
\begin{equation}\label{eq:t_eq}
t_{eq} = t_0 (R_c H_0)^{1+\frac{2\bar w}{\delta(1+\bar w)}}
\end{equation}
Finally, we define a time $t_{dom}$, at which the dangerous modes 
formally
dominate
the universe, corresponding to $t^2 \rho(t) \sim 1$.  Then one finds,
\begin{equation}
t_{dom} = t_{eq} \left( \left[ t^2 \rho(t) \right]_0 \right)^{1/\delta}.
\end{equation}
Chaos is controlled provided that the dangerous modes do not dominate
before a Planck time from the big crunch, corresponding to 
$t_{dom} < t_{PL}$.

Having established the formulae we will need for our estimate, we may now
insert reasonable values for our variables. Let us assume the contraction phase 
begins when the Hubble parameter  is of order the present value 
$H_0$ (as occurs in ekpyrotic and cyclic models).  Then,  $1/H_0 \sim 10^{61}$ 
$L_{PL}$.  As a first example, we assume that
$\bar w = 1$ during the contraction as is characteristic 
of compactifications of Kasner universes.
  A typical value for $\delta$ that arises from working with the $p$--form
spectrum of string models and gravity is $\delta = 2$, although the precise 
value of 
$\delta$
will of course depend on the specific model under consideration.  If we take
$R_c \sim 10 \cdot L_{PL}$, as an example, then we find the suppression factor
\er{eq:rhoAtHisM} at $H \sim m$ to be,
\begin{equation}
(R_c H_0)^{\frac{2\bar w}{1+\bar w}} \sim 10^{-60}.
\end{equation}
The dangerous modes grow to have the same fractional 
energy density as they had at the beginning  of the contracting phase at the
time,
\begin{equation}
t_{eq}  \sim 10^{-29} \; t_{PL}.
\end{equation}
Thus, the dangerous modes cannot grow to be even as relevant as at the beginning 
of 
cosmic contraction, until the universe is well within the quantum 
regime.  We would need to approach even closer to the big crunch for these modes
to dominate the universe, but at this point we no longer expect our classical 
equations
to be valid. 

As another example, we consider a case with large compact dimensions, where
$m$ is of order the weak scale, $10^{-16}$ $M_{PL}$, which
corresponds to taking $R_c \sim 10^{16}$ $L_{PL}$.  Now we find that the suppression
when the Hubble radius equals the compactification scale is,
\begin{equation}
(R_c H_0)^{\frac{2\bar w}{1+\bar w}} \sim 10^{-45},
\end{equation}
and,
\begin{equation}
t_{eq} = 10^{-6} \; t_{PL}.
\end{equation}
Again, we will be well within the quantum gravity regime before the dangerous 
modes
can potentially dominate.
If  $\bar{w}\gg 1$ during the contraction phase, as occurs in ekpyrotic and 
cyclic models,  the dangerous modes are suppressed by a much greater factor than 
in these examples, as is evident from the expressions above. 

As a final relevant issue, note that we have taken the mass $m$ of the dangerous 
modes to be constant in time.
Generally we expect that the mass $m$ of a given field will be time dependent.
This occurs since the field's mass is determined by the compactification 
manifold $\isC$, and $\isC$ will be evolving with time during cosmic 
contraction.
Let us say that the 
mass $m(t)$ evolves as,
\begin{equation}
m(t) = m_0 \left(\frac{t}{t_0}\right)^b,
\end{equation}
during the contracting phase, where for simplicity we take $\bar w=1$.  Then, 
because the energy density will go like $\rho \sim m/a^3$, we find that
\begin{equation}
t^2 \rho = \left[t^2 \rho\right]_0 \left(\frac{t}{t_0}\right)^{1+b}
\end{equation}
However, the time $t_1$ at which $m/H \sim 1$ is now,
\begin{equation}
t_1^{1+b}  = \frac{t_0^b}{m_0},
\end{equation}
and therefore the suppression when $m/H \sim 1$ is,
\begin{equation}
\frac{ \left[t^2 \rho\right]_1 }{ \left[t^2 \rho\right]_0 }
= (H_0/m_0) \sim (H_0 R_{c,0}),
\end{equation}
where $R_{c,0}$ is the characteristic length of $\isC$ when 
$t=t_0$.
This is precisely the factor found in the case where $m$ is constant in 
time \er{eq:rhoAtHisM}.  
The difference is that the time $t_1$, at which $m/H \sim 1$, shifts.
This shift compensates for the different growth rate of $\rho$, with the
net result that the suppression factor remains the same.

From the equations above, it is clear that there is a potential problem
if $b < -1$, which would lead to $t^2 \rho$ increasing during the
contracting phase.  However, our assumption that the universe is of
Kasner type excludes this possibility.  Field masses $m$ are related to
the compactification length scale $R_c$ by $m \sim 1/R_c$.  In a Kasner
universe, we expect that $R_c \sim t^p$, with $p$ a Kasner exponent, or
average of several Kasner exponents.  However, the Kasner conditions 
\er{eq:kasnerCond} or \er{eq:kasnerScalarCond}
imply that $p \le 1$, and thus $m$ must vary more slowly than $1/t$.
This implies that $b \ge -1$, and so the suppression operates as before.

\subsection{The $p$--form spectrum}\label{ss:pFormSpectrum}

Having established that the massless modes are the only relevant modes near
the big crunch, we now describe how these massless modes are determined in terms
of the compactification manifold $\isC$.  The story is familiar
from the study of higher--dimensional models of particle physics
\cite{Pol98,GSW87,Duf86}
although we discuss
some special features of the time dependent situation which must be taken into 
account.
An useful feature of the $p$--form mass spectrum is that the existence of 
massless
modes is determined entirely by the topology of $\isC$, and not by its metric.
This simplifies the task of finding manifolds $\isC$ that lead to controlled
chaos, since we need only specify their topological properties.

\subsubsection{Time independent compactification}

First, we will review the situation for the 
time independent case.  For clarity, we will neglect here
 the exponential coupling to the 
dilaton field,
which amounts to an overall multiplication by $e^{\lambda \phi}$ of the 
Lagrangian
density.  The coupling is fully accounted for in the analysis given in Appendix 
1.
The action for a $p$--form gauge potential $A_p$ in $4+n$ dimensions is, 
\begin{equation}\label{eq:pFormAction}
S = - \frac{1}{(p+1)!} \int  \diff A_p\cdot \diff A_p  \, \sqrt{-G} \; d^{4+n}x 
,
\end{equation}
where $A_p$ can depend on all coordinates, and has indices along both 
$\nonC$ and $\isC$,
\begin{equation}
A_p = \left[ A_p ( x^\mu, x^m ) \right]_{\alpha_1 \alpha_2 \dots \alpha_r a_1 
a_2
\dots a_{p-r} } . 
\end{equation}
The conventional compactification analysis begins with an expansion of the 
$p$--form $A_p$ as,
\begin{equation}\label{eq:pFormExpansion}
A_p = \sum_{r+s=p} \sum_i \alpha^{(i)}_r \wedge \beta^{(i)}_s
\end{equation}
using a basis of $s$--forms $\beta^{(i)}_s$,
with indices and coordinate dependence only along $\isC$, 
\begin{equation}
\beta^{(i)}_s = \left[ \beta^{(i)}_s(x^m) \right]_{a_1 a_2 \dots a_s}.
\end{equation}
The abstract index $i$ labels the $s$--form under
consideration, and when $\isC$ is compact it takes discrete values, infinite in 
number.
The ``coefficients" in this expansion are 
$r$--forms $\alpha^{(i)}_r$, that depend on the noncompact coordinates on
$\nonC$, and have indices along $\nonC$ only,
\begin{equation}
\alpha^{(i)}_r = \left[ \alpha^{(i)}_r(x^\mu) \right]_{\alpha_1 \alpha_2 \dots 
\alpha_r}.
\end{equation}
It is convenient to choose the gauge $\difs \beta_s^{(i)} = 0$, and select the
$\beta^{(i)}_s$ to be eigenfunctions of the Hodge--de Rham Laplacian
$\Delta = \diff \difs + \difs \diff $ on $\isC$, with eigenvalues 
$\lambda^{(i)}_s$,
\begin{equation}\label{eq:eigenForms}
\Delta \beta^{(i)}_s = \lambda^{(i)}_s \beta^{(i)}_s.
\end{equation}
The $\beta^{(i)}_s$ are normalized so that,
\begin{equation}\label{eq:staticNorm}
\int_{\isC} \beta^{(i)}_s \cdot \beta^{(j)}_s \, \sqrt{-f} \; d^n x =  
\delta_{ij}.
\end{equation}
Substituting the expansion \er{eq:pFormExpansion} into the original action
\er{eq:pFormAction} results in,
\begin{equation}\
S = - \sum_{i,s}  \frac{1}{(r+1)!} \int \left(
 \diff \alpha^{(i)}_r \cdot \diff \alpha^{(i)}_r  
  +  \lambda^{(i)}_{s} \alpha^{(i)}_r \cdot \alpha^{(i)}_r \right) \, \sqrt{-h} 
\;
  d^{4}x,
\end{equation}
where we have  rescaled the $\alpha^{(i)}_r$ by a constant
in order to canonically normalize
the kinetic terms.  This demonstrates that a single $p$--form in $4+n$ 
dimensions
yields many $r$--forms $\alpha^{(i)}_r$ after compactification, whose masses 
$m^{(i)}_r$ are related to
the  eigenvalues $\lambda^{(i)}_s$ by $\left( m^{(i)}_r \right)^2 = 
\lambda^{(i)}_s$.
These are the ``Kaluza--Klein" modes.
The operator $\Delta $ has a positive semi--definite spectrum
on manifolds which, like $\isC$, have a Euclidean metric.  Therefore the 
effective masses
are all real.

The $(p-s)$--forms with zero effective mass are determined 
entirely by the topology  of $\isC$, and not by its metric structure.
As discussed above, massless $(p-s)$--forms arise from $s$--forms $\beta_s$ 
satisfying
$\Delta \beta^{(i)}_s = 0$, conventionally termed ``harmonic" forms.
Hodge's theorem \cite{Nak90,EGH80} states that
the number of harmonic $s$--forms is equal to the dimension of $\coH s \isC $,
 the $s^{th}$ de Rham cohomology class of $\isC$.  
The dimension $\dim \coH s \isC$ is also known as the $s^{th}$ Betti number of
$\isC$, conventionally denoted $b_s(\isC)$.
This is a topological invariant, which does not change under smooth 
deformations of $\isC$ and its associated metric structure.
The Poincar\'{e} duality theorem
\cite{Mas91,Mad97} gives a simple geometric interpretation of these cohomology
classes; the quantity $\dim \coH{ s }{\isC}$ counts the number of 
$s$--dimensional submanifolds that can ``wrap" $\isC$, and cannot
be smoothly contracted to zero.  
In this counting, two submanifolds are considered equivalent if one can be 
smoothly
deformed into the other.
Thus, for every inequivalent noncontractible submanifold of $\isC$ with
dimension $s$, a $p$--form gives rise to a massless $(p-s)$--form field after
compactification.

\subsubsection{Time dependent compactification}

The case where the compactification manifold $\isC$ changes with time introduces 
new 
features, but in the end does not substantially modify the conclusions reached 
above.
The main difference is that it is no longer possible to assume that the eigenbasis 
of 
forms $\beta^{(i)}_s$ defined by (\ref{eq:eigenForms}) depends only on the 
compact
 coordinates $x^m$.  In particular, the
forms will depend on time.  This introduces additional cross terms which must be 
taken 
into account.  Below, we will neglect the variation of $\isC$ along directions 
$x^M$ 
other than time.
We  denote by $\diff |_\isC$ the exterior derivative tangent to the
manifold $\isC$.  Thus,
\begin{equation}
\diff \beta^{(j)}_s = \diff |_\isC \beta^{(i)}_s + \diff t \wedge \dot 
\beta^{(i)}_s.
\end{equation}
We may now use our freedom to choose the basis modes $\beta^{(i)}_s$, and 
define the modes $\beta^{(i)}_s(t)$ with $\lambda^{(i)}_s(t) > 0$
to be the instantaneous eigenforms of the 
Hodge--de Rham operator, restricted to act on $\isC$ only,
\begin{equation}
\Delta |_\isC \beta^{(i)}_s(t) = \lambda^{(i)}_s(t) \beta^{(i)}_s(t).
\end{equation}
We find it convenient to
 relax the requirement that the zero modes $\beta^{(i)}_s(t)$ with
$\lambda^{(i)}_s = 0$ be eigenforms
of the Hodge--de Rham Laplacian.  Instead, we will merely require that they be
representatives of the de Rham cohomology of $\isC$.  Inspection of the
reduced action shows that this condition is sufficient to guarantee that the
zero modes still result in massless form fields.
Furthermore, we adopt the normalization convention,
\begin{equation}\label{eq:timeDnorm}
\int_{\isC} \beta^{(i)}_s \cdot \beta^{(j)}_s \, \sqrt{-f}\; d^n x = 
\delta_{ij} \int  \, \sqrt{-f}\; d^n x .
\end{equation}
This differs from the usual normalization convention  (\ref{eq:staticNorm})
by only a multiplicative constant in the static case.  It has the advantage of 
not 
introducing
any spurious time dependence of the $\beta^{(i)}_s$ from the changing volume of 
$\isC$.
Maintaining this normalization condition requires,
\begin{equation}\label{eq:timeDnorm2}
\int_{\isC} \dot \beta^{(i)}_s \cdot \beta^{(j)}_s \, \sqrt{-f}\; d^n x = 0 .
\end{equation}
With these conventions and definitions, we find that the $p$--form action 
(\ref{eq:pFormAction}) splits into two parts, $S_1$ and $S_2$, with
\begin{equation}
S_1 =  - \sum_{s,i} \frac{1}{(r+1)!} \int  \diff \alpha^{(i)}_r \cdot \diff   
  \alpha^{(i)}_r  
  +  \lambda^{(i)}_s (t) \alpha^{(i)}_r \cdot \alpha^{(i)}_r 
\, \sqrt{-G} \; d^{4+n}x
\end{equation}
and,
\begin{equation}
S_2 = - \sum_{s,i} \frac{2}{(r+1)!} \int (\alpha^{(i)}_r \wedge {\dot 
\beta}^{(i)}_s)
  \cdot ({\dot \alpha}{(i)}_r \wedge \beta^{(i)}_s )
  + (\alpha^{(i)}_r \cdot \alpha^{(i)}_r) ({\dot \beta}^{(i)}_s \cdot 
  {\dot \beta}^{(i)}_s) \, \sqrt{-G} \; d^{4+n} x,
\end{equation}
where terms that are identically zero due to mismatched indices are not 
included.
Again, we have rescaled  the $\alpha^{(i)}_r$ to obtain canonically normalized 
kinetic terms.
The terms in $S_2$ threaten to substantially modify the action in the time 
dependent
case.  However, these terms vanish or are negligible. 
The first term is zero due to our normalization convention 
(\ref{eq:timeDnorm}) and its consequence (\ref{eq:timeDnorm2}).  
The second term is a contribution to the effective mass of
$\alpha^{(i)}_r$.  
For the $\lambda^{(i)}_s = 0$ modes, the representatives $\beta_s^{(i)}$
of the de Rham cohomology are time--independent, and so the $\dot \beta^2$ terms 
vanish.
When $\lambda^{(i)}_s > 0$, the additional contribution to the effective mass 
will be 
positive, but since these modes are already massive it will not change the 
qualitative
features of their behavior.
Thus, time dependent compactifications do not substantially modify the $p$--form 
spectrum;
massless modes are still given by the de Rham cohomology of $\isC$.

\subsection{The gravitational spectrum} \label{ss:gravitySpectrum}

A key property of Kaluza--Klein reduction is that degrees of freedom in
the full metric $G_{MN}$ appear in lower dimensions as metric, vector, and 
scalar
degrees of freedom.  As in the $p$--form compactification discussed above,
the masses of these fields depend on the properties of the compactification 
manifold
$\isC$.  In contrast to the $p$--form case, the masses are not determined by the
cohomology of $\isC$, but by the existence of Killing fields on $\isC$.
The properties of  Kaluza--Klein reduction, along with our discussion of chaos
in Section \ref{s:review}, provide some useful simplifications.
Since some of the metric degrees of freedom in the higher dimensional theory 
appear as
zero-- and one--forms, chaos arising from these degrees of freedom can be
suppressed if they acquire a mass, just as in the conventional $p$--form case.

As an explicit example of the reduction process, and of how chaos in higher and 
lower
dimensions are related, we consider below the simple case with a single extra 
dimension
\cite{Bel73}. The Kaluza--Klein reduction begins with a reparameterization of 
the metric,

\begin{equation}\label{eq:KKreparam}
G_{MN} =
\begin{pmatrix}
e^{-q\phi} g_{\mu\nu}+ e^{2q\phi} A_\mu A_\nu &  e^{2q\phi} A_\nu \\
 e^{2q\phi} A_\mu & e^{2q\phi}
\end{pmatrix},
\end{equation}
where we assume that $A_\mu$ and $\phi$ are independent of the fifth dimension.  
We substitute this metric into the Einstein--Hilbert
action and integrate over the fifth dimension.
The coefficient $q = \sqrt{2/3}$ is chosen so that
the scalar field $\phi$ has a canonically normalized kinetic term in the 
resulting
action,
\begin{equation}\label{eq:5dKK}
S = \int R(g) - (\partial \phi)^2 - \frac{1}{4} e^{\sqrt{6} \phi} F_{\mu\nu} 
F^{\mu \nu}
\, \sqrt{-g} \; d^4 x ,
\end{equation}
with $F_{\mu\nu} = \partial_\mu A_\nu - \partial_\nu A_\mu$.
This describes Einstein gravity coupled to a scalar field $\phi$, and a vector 
field with
$\phi$--dependent coupling.  It can be seen that all of the five--dimensional 
metric
degrees of freedom in \er{eq:KKreparam} are reproduced in this 
four--dimensional action.  Furthermore, the vector term in the action possesses 
an
 exponential
coupling to $\phi$, of the type introduced in \er{eq:pFormDef}. 
Our starting point, the five dimensional pure gravity theory, is chaotic since 
the
gravitational stability conditions cannot be satisfied for any choice of the 
Kasner
exponents.  After reduction, chaos also inevitably arises since the 
gravitational and one--form stability conditions cannot be satisfied 
simultaneously.
Thus violations of the gravitational stability conditions in five dimensions can 
appear
as violations of the $p$--form stability conditions in four dimensions.
As we will discuss in more detail below, the preservation of chaos is not a 
generic 
feature of Kaluza--Klein reduction in dimensions greater than one.

The example above is limited to a single extra dimension, and neglects metric
modes that depend on the fifth coordinate.  Below, we consider the general case, 
and
calculate the effective masses of all Kaluza--Klein vector fields
with an arbitrary number of extra dimensions.  We will find that these masses 
are
zero only when $\isC$ possesses Killing vectors.
This calculation parallels standard treatments of Kaluza--Klein reduction
\cite{Duf86}, but in these treatments the fact that $\isC$ may not
possess isometries is often not emphasized.
For $n > 1$ extra dimensions, we will generalize the decomposition
\er{eq:KKreparam} using the  vielbein formalism.
We begin by defining one form fields $e^A = {e_M}^A {\diff x}^M$ so that,
\begin{equation}
ds^2 = {e}^A {e}^B \eta_{AB},
\end{equation}
with $\eta_{AB}$ the $(4+n)$--dimensional Minkowski metric.  The ${e_M}^A$ are 
chosen so
that,
\begin{subequations}
\begin{align}
h_{\mu\nu} &= {e_\mu}^\alpha  {e_\nu}^\beta \eta_{\alpha\beta},\\
f_{mn}     &= {e_m}^a {e_n}^b \delta_{ab},\\
{e_\mu}^a  &= K^{aj}(x^m) A^j_\mu(x^\mu),\\
{e_m}^\alpha &= 0,
\end{align}
\end{subequations}
where $\delta_{ab}$ the Euclidean flat space metric.
The $K^{aj}$ are a basis for vector fields on $\isC$, indexed by $j$, that 
depend
only on the compact coordinates $x^m$. 
The coefficients in this expansion are the
$A^j_\mu$, which depend only the noncompact coordinates $x^\mu$.  
The $A^j_\mu$, known as Kaluza--Klein vectors, will emerge after 
compactification
as vector fields on the noncompact space $\nonC$.
The commutators
of the $K^{aj}$ define a set of structure constants 
${f^{jk}}_{l}$,
\begin{equation}\label{eq:structureConstants}
[ K^{j}, K^{k} ] = {f^{jk}}_{l} K^l .
\end{equation}
The calculation is most conveniently carried out using an orthonormal 
basis $\{ {\hat e}^a, {\hat e}^\alpha \}$ given by,
\begin{equation}
{\hat e}^\alpha = e^\alpha \qquad {\hat e}^a = e^a - K^{aj} A^j_\mu {\diff 
x}^\mu,
\end{equation}
in which the line element assumes the simple form
$ds^2 = {\hat e}^\alpha {\hat e}^\beta \eta_{\alpha\beta} + 
{\hat e}^a {\hat e}^b \delta_{ab}$.
In the event that some of the $K^{aj}$ are Killing fields on $\isC$, 
then the lower dimensional
theory will possess a gauge symmetry.  The Killing fields are generators of the 
isometry group of $\isC$, and this isometry group reemerges as the gauge group 
in the
lower dimensional theory.  This motivates the definition of a
``field strength" $F^i_{\mu\nu}$ as,
\begin{equation}
F^i = \diff A^i + \frac{1}{2} f^{ijk} A^j \wedge A^k.
\end{equation}
In the general case, the Killing fields alone do not provide a full basis for 
vector
fields on $\isC$.  Thus, in addition to the massless modes (if any) of the gauge 
theory,
there will also be an infinite set of massive gauge fields in the lower 
dimensional
theory.

In order to derive the mass spectrum in the lower dimension explicitly, we may
use the vielbeins to decompose the gravitational action in $4+n$ dimensions.
The spin connections are, 
\begin{subequations}
\begin{align}
{\hat \omega}_{ab} &= \omega_{ab} - \frac{1}{2} ( \nabla_b K_a^j - 
\nabla_a K_b^j ) A^j_\beta e^\beta, \\
{\hat \omega}_{a\beta} &= \frac{1}{2} K_a^j F^j_{\alpha\beta} e^\alpha
+ \nabla_{(b} K_{a)}^j A^j_\beta {\hat e}^\beta, \\
{\hat \omega}_{\alpha\beta} &= \omega_{\alpha\beta} + \frac{1}{2} K_b^j 
F^j_{\alpha\beta}
{\hat e}^b,
\end{align}
\end{subequations}
where $\omega_{ab}$ and $\omega_{\alpha\beta}$ are the spin connections defined
by the metrics $f_{mn}$ on $\isC$,
and $h_{\mu\nu}$ on $\nonC$, respectively.  Using these spin connections to 
compute
the Ricci scalar, one obtains,
\begin{equation}\label{eq:factoredR}
R(G) = R(h) + R(f) - \frac{1}{4} K^j_a K^{ka} F^j_{\mu\nu} F^{k \mu\nu}
- 2 \nabla_{(c} K^j_{d)} \nabla^{(c} K^{kd)} A^{j\mu} A^k_\mu.
\end{equation}
We see that a mass term for the  $A^j$ has appeared. 
Upon integrating over the compact coordinates, one arrives at the Jordan frame 
action,
\begin{equation}\label{eq:jordanAction}
S = \int  \left( W(f) R(h) - S(f)
-\frac{1}{4} \alpha_{jk} F^j_{\mu\nu} F^{k \mu\nu}
- \beta_{jk} A^{j\mu} A^k_\mu \right) \; \sqrt{h} \, d^{4}x.
\end{equation}
where,
\begin{subequations}\label{eq:jordanActionBits}
\begin{align}
W(f)        & = \int_\isC  \; \sqrt{f} \, d^n x \\
S(f)        & = - \int_\isC R(f) \; \sqrt{f} \, d^n x \\
\alpha_{jk} &= \int_\isC K_{ja} K_k^a \; \sqrt{f} \, d^n x \\
\beta_{jk}  &= 2 \int_\isC \nabla_{(c} K_{jd)} \nabla^{(c} K_{k}^{d)}\; 
               \sqrt{f} \, d^n x .
\end{align}
\end{subequations}
The $W(f)$ factor may be removed by a rescaling of the metric $h_{\mu\nu}$, 
putting
the action in the Einstein frame form.
The $S(f)$ term will yield a system of scalar fields, of which $\phi$ in our 
five dimensional reduction \er{eq:5dKK} is an example.
Applying the Gram--Schmidt orthonormalization process to the $K^j$, one can 
reduce 
$\alpha_{jk} = \delta_{jk}$, giving the vectors a canonical kinetic term.  
Thus, we see Kaluza--Klein reduction of pure gravity results in a
 theory with scalars and vector fields, generalizing
the $n=1$ result discussed above.

Of crucial importance to the present work is that massless Kaluza--Klein vectors
are in one--to--one correspondence with Killing fields on $\isC$, or 
equivalently
the zero eigenvalues of $\beta_{jk}$.
This follows from the fact that $\beta_{jk}$ functions as a mass matrix
for the Kaluza--Klein vector fields.  Since $\beta_{jk}$ is symmetric, 
we are guaranteed that $m^2$ will be real for all modes.
In our discussion of $p$--form fields, we were able to apply powerful results 
regarding the
Hodge--de Rham operator $\Delta$ that guaranteed that $m^2 \ge 0$, regardless of 
the 
topology and metric structure of $\isC$.
In the present situation, we have no guarantee that the masses of Kaluza--Klein 
vectors will satisfy $m^2 \ge 0$, or equivalently that the eigenvalues of the 
mass 
matrix are nonnegative.  

In this work we will assume that all eigenvalues of the mass matrix
are nonnegative, so that $m^2 \ge 0$ for all
Kaluza--Klein vectors.  In the general case, 
it is necessary to compute
$\alpha_{jk}$ and
$\beta_{jk}$ for each manifold of interest, and then check that this
assumption holds on a case--by--case basis.  A simple example is provided by
the $n$--torus $\torus n$.  Realizing the torus as $\real n / \Zto n$, with
coordinates $(\theta_1 \dots \theta_n)$ ranging on $(0,2\pi)$, a 
convenient basis
for vector fields on $\torus n$ is provided by,
\begin{equation}
K^{(a,n)} = 
\frac{\sqrt{2} \hat \theta^a}{(2\pi)^{n/2}} \times
\begin{cases}
  \cos n\theta , & \qquad \mbox{for } n \ge 0, \\
  \sin n\theta , & \qquad \mbox{for } n < 0
\end{cases}
\end{equation}
where $n \in \Zto{}$, the $\hat \theta^a$ are unit vectors associated to
each coordinate, and
 $(a,n)$ label each basis field, replacing the abstract indices
used above.   Substituting
this into 
\er{eq:jordanActionBits} we find,
\begin{subequations}
\begin{align}
\alpha_{(a,n) (b,m)} &= \delta_{ab} \delta_{nm}, \\
\beta_{(a,n) (b,m)} &= 2n^2 \delta_{ab} \delta_{nm}.
\end{align}
\end{subequations}
The Kaluza--Klein vector fields are therefore canonically normalized,
with masses $\sqrt{2} n$ for $n=0,1,2 \dots$, showing our assumption is valid
in this case.
More sophisticated examples may be found in the literature \cite{Duf86}. 
As we will explain in more detail below, the $m^2 \ge 0$ assumption will
enable us to treat $p$--forms and Kaluza--Klein vectors on the same footing.

\section{Selection Rules for the Stability Conditions}\label{s:msc}

With the tools developed in the previous sections, we are now prepared to 
discuss conditions on $\isC$ that result in controlled chaos.  We will show
that the gravitational, electric and magnetic stability conditions,
introduced in Section \ref{s:review}, are modified by compactification.  
Not all of the stability conditions remain relevant, 
and only a subset need be satisfied to ensure that chaos is 
controlled.  This subset is defined by the ``selection rules" that are the focus 
of this 
section.  The selection rules that determine when a stability condition remains 
relevant 
are given for matter fields in Section \ref{ss:matterMSC}, and for gravitational
modes in \ref{ss:gravityMSC}.  
The selection rules, in turn, are determined by the de Rham cohomology (in the
$p$--form case) and existence of Killing vectors (in gravitational case) of the
compactification manifold $\isC$.
Here, we focus on discussing the origin of the selection rules. 
 Once established, we will use them  
to find compactifications that control chaos in Section \ref{s:examples}.

\subsection{The $p$--form selection rules}\label{ss:matterMSC}

The discussion in Section \ref{s:cc} enables us to 
define  selection rules for the electric and magnetic
stability conditions.  Each component of the $p$--form field results in an 
electric
or magnetic stability condition, which expresses whether the energy density in 
that
component scales rapidly enough to dominate the energy density of the universe 
and
cause chaos.  If this component gains a mass by compactification, then we have
shown that it scales too slowly to be cosmologically relevant, and therefore we
should ignore the corresponding stability condition.
 Thus we should ignore all 
electric and magnetic stability conditions involving indices that do not 
correspond to massless $p$--form modes.  This results in the following selection 
rule,
 
\begin{quote}
\textbf{The $p$--form Selection Rule:}  
When $\dim\coH{s}{\isC} = 0$ for some $s$, ignore the subset of 
 $p$--form stability conditions,
\begin{subequations}\label{eq:EBstability2}
  \begin{align}
    \sum_{p}   p_j - \frac{1}{2} \lambda p_\phi &> 0 \qquad \mbox{(electric)}, 
\\
    \sum_{p+1} p_j - \frac{1}{2} \lambda p_\phi &< 1 \qquad \mbox{(magnetic)},
  \end{align}
\end{subequations}
with $s$ Kasner exponents along the compact space $\isC$.   Retain only those
stability conditions with $s$ exponents along $\isC$ and $\dim\coH{s}{\isC} \ne 
0$.
\end{quote}
 
\noindent In this section, we will use the results of previous sections
to prove this rule, and give some simple examples of its use.

This rule arises from considering the $p$--form modes that give rise to 
massless fields after compactification. 
A $p$--form $A_p$ gives rise to a massless $r$--form $\alpha_r$
if and only if  $A_p$ is of the form,
\begin{equation}
A_{p} = \alpha_r \wedge \beta_s
\end{equation}
where $\beta_s \in \coH s \isC$, and $r+s = p$.  Since
$\diff \beta_s = 0$, this gauge potential results in the 
field strength,
\begin{equation}\label{eq:masslessPFormFieldStrength}
F_{p+1} = (\diff \alpha )_{r+1} \wedge \beta_s.
\end{equation}
When the energy density of this field is calculated, one finds  a
stability condition involving exactly $s$ Kasner exponents along $\isC$, 
and the remainder
along $\nonC$.  Since only stability conditions of this type correspond to 
massless
modes, they are the only ones that should be retained.  

The case
\er{eq:masslessPFormFieldStrength} deals only with field strengths 
having at least one index along the noncompact space $\nonC$.  In fact,
the same selection rule applies when all indices of the field strength
are along $\isC$.  In this case, the field strength must satisfy the Bianchi
identity and the Gauss law,
\begin{equation}\label{eq:nonZeroModes}
\diff F_{p+1} = 0, \qquad \difs F_{p+1} = 0.
\end{equation}
Field strengths of this type are commonly termed ``nonzero modes" \cite{GSW87}.  
The conditions \er{eq:nonZeroModes} imply that $F_{p+1}$
is harmonic, and by Hodge's theorem  the number of such forms is given by
$\dim \coH {p+1} \isC$.  
When $\dim \coH {p+1} \isC$ vanishes, we cannot have a $p$--form field
strength with all indices along $\isC$, and we should therefore delete the 
corresponding
stability condition, with $(p+1)$ Kasner exponents  along $\isC$.  Thus this 
case
falls under the $p$--form selection rule as well.

The selection rule may be illustrated by comparing compactification on a sphere 
$\sphere n$ and a torus $\torus n$ .  These manifolds encompass
 the  best and worst case scenarios for controlling chaos through
compactification.  
The sphere has the minimum number of massless modes for any 
orientable compact 
manifold, while the torus has massless modes for every dimension and
involving every combination of indices on $\isC$.  
Therefore compactification on $\torus n$ and $\sphere n$ will have very 
different
influences on chaotic behavior.

Compactification on $\torus n$ does not modify any of the $p$--form stability 
conditions.
The cohomology classes of the torus are,
\begin{equation}
\dim \coH r {\torus n} = \frac{n!}{r! (n-r)! },
\qquad 0 \le r \le n.
\end{equation}
If we realize the torus as $\real n / \Zto n$, with coordinates
$\theta_1, \dots \theta_n$, then we may choose the following 
set of generators for the $r^{th}$ de Rham class,
\begin{equation}
\omega_r = \diff \theta_{j_1} \wedge \diff \theta_{j_2} \wedge \dots \wedge 
\diff 
\theta_{j_r},
\end{equation}
where $\{j_r\}$ are any set of distinct indices on $\torus n$. Therefore,
massless modes exist for $p$--form fields with any combination of indices
along the $\torus n$.  Any $p$--form stability condition that appears
in the noncompactified theory will remain in the compactified theory.

By contrast, compactification on a sphere $\sphere n$, with $n > 1$, 
deletes many of the stability conditions.  
The sphere has
only two nonzero cohomology groups, each of unit dimension;  
\begin{equation}
\dim \coH{r}{ \sphere{n}} = 
\begin{cases}
1, \qquad &\text{for} \quad r = 0 \quad \text{and} \quad r=n \\
0, \qquad &\text{otherwise}. 
\end{cases}
\end{equation}
The class $\coH{0}{\sphere n}$ is generated by the constant scalar function
on $\sphere n$, while the class $\coH{n}{\sphere n}$ is generated by the
volume form,
\begin{equation}
\omega_n = 
\sqrt{g}\, \diff \theta_1 \wedge \diff \theta_2  \wedge \dots \wedge \diff 
\theta_n ,
\end{equation}
where $g_{mn}$ is the metric, $g = \det g_{mn}$, and $\theta_j$ 
the coordinates on the sphere.
Massless modes therefore 
contain either no indices along the $\sphere n$, or all indices at
once.  This implies that the only surviving stability conditions are those with 
either no
internal Kasner exponents, or all internal Kasner exponents together.
In the case where $n>p$, none of the internal Kasner exponents appear at all, 
and
only those stability conditions involving Kasner exponents on $\nonC$ survive.

Our statement of the selection rule is the strongest possible in the generic 
case, 
and fortunately
also the most conservative in terms of deleting the minimum number of stability
conditions.
Manifolds with a specific relationship between the
frames $\sigma^a$ appearing in \er{eq:genKasner} and the cohomology 
representatives
of $\isC$ may require that we delete additional $p$--form stability conditions.
For example, consider the case in which $\isC$ factors as
 $ \isC = \mc M_1 \times \mc M_2$, both 
topologically and metrically.  A straightforward application of the 
selection rules results in retaining all
stability conditions with $s$ indices along $\isC$ whenever 
$\dim \coH{s}{\isC} \ne 0$.  
These stability conditions correspond to $p$--form modes with $s_1$ indices 
along
$\mc M_1$ and $s_2$ indices along $\mc M_2$, with $s_1 + s_2 = s$.
However, in general a massless mode will not exist for every choice of
$s_1$ and $s_2$, and therefore we may be able to delete
additional stability conditions.  We should only retain the even smaller subset
of stability conditions with $s_1$ Kasner indices along $\mc M_1$ and
$s_2$ indices along $\mc M_2$ when $\dim \coH{s_1}{\mc M_1} \ne 0$ and
$\dim \coH{s_2}{\mc M_2} \ne 0$.
Generally, however, we do not expect any special relationship between the
$\sigma_j$ and the cohomology classes.  In this example, we have imposed the 
condition by hand
that the $\sigma_j$ point only along exactly one of $\mc M_1$ or $\mc M_2$.
Examples such as this 
one must be considered on a case--by--case basis, and lie beyond the scope of 
our 
selection rule.

\subsection{The gravitational selection rules}\label{ss:gravityMSC}

The selection rules for the gravitational stability conditions arise in a manner
similar to those for the $p$--form modes. 
We identify the degree of freedom corresponding to each stability condition,
and then ignore the stability condition if the degree of freedom gains a
mass through compactification.  Unlike the $p$--form case, in general
one must also check that the masses gained in this way satisfy
$m^2 \ge 0$, as discussed 
at the end of Section \ref{ss:gravitySpectrum}.
In this way, we arrive at a selection
rule for the gravitational stability conditions,

\begin{quote}
\textbf{The Gravitational Selection Rule:}  When $\isC$ possesses no Killing
vectors, retain only the subset of gravitational stability conditions, 
\begin{equation}\label{eq:gravStability2}
p_i + p_j - p_k < 1, \qquad \mbox{all triples} \quad i,j,k ,
\end{equation}
with all three Kasner exponents along
$\nonC$, or all three along $\isC$, and ignore stability conditions with a 
mixture of
exponents
along both $\nonC$ and $\isC$.
\end{quote}

\noindent In proceeding, we are guided 
by the Kaluza--Klein reduced Jordan frame action \er{eq:jordanAction}.
  Clearly, gravitational stability conditions involving
three Kasner exponents along the $\nonC$ should be retained, as the 
corresponding
modes do not
gain a mass from compactification.  These represent the metric degrees of 
freedom
in the lower dimensional theory.  Stability conditions involving three 
Kasner exponents along the compact direction should also be retained.  
Physically,
these correspond to metric degrees of freedom on compact space $\isC$.  While 
these
appear as scalar fields in the lower dimensional theory, they can result in a 
subtle 
form of chaos.  Violations of these stability conditions appears as a 
chaotic system of interacting scalars in the lower dimensional theory.  Thus,
to ensure that all degrees of freedom are evolving smoothly to the big crunch, 
we
should retain these stability conditions.

Compactification can delete the ``mixed" stability conditions, those with 
Kasner exponents along both $\nonC$ and $\isC$.  These appear as the
kinetic and mass terms for the Kaluza--Klein vectors in \er{eq:jordanAction}.
When the compact space $\isC$ possesses no Killing vectors, then these vector 
fields
acquire a mass and become cosmologically irrelevant.  When $\isC$ possesses even
one Killing vector, then in general none of the mixed stability conditions can 
be
discarded.  This is because a single Killing field will generally involve all 
indices 
along $\isC$, and also results in a vector field with arbitrary indices on 
$\nonC$.  
As in the $p$--form case, there can be special cases where additional stability
conditions may be deleted.  However, as we are more interested in the generic 
case
we will not discuss examples here.

\section{Examples}\label{s:examples}

The previous sections have established that compactification allows us to ignore
a number of the gravitational and $p$--form stability conditions.  At this 
point,
it is natural to ask if there are examples where enough stability conditions
are deleted to control chaos.  We show below that this is indeed the case, by 
giving
explicit examples from both pure Einstein gravity and the low--energy bosonic 
sectors
of string theory.  We will discuss several solutions with controlled chaos in 
string
models with $\mc N =1$ supersymmetry in ten dimensions, and will show how these 
solutions
are interrelated by string duality relationships.

For simplicity, we focus on regions on the Kasner sphere near what we will term
``doubly isotropic" solutions.  These are solutions in which $a$ Kasner 
exponents
take the value $p_a$, and $b$ take the value $p_b$, with $a+b = D-1$.  
In the absence of a dilaton, the Kasner conditions \er{eq:kasnerCond}
result in a quadratic equation for $p_a$ and $p_b$, and therefore two solutions
for each choice of $a$ and $b$.  When a dilaton is present, then there
are two one--parameter families of $p_a$ and $p_b$, which depend on the value of 
the
dilaton ``Kasner exponent" $p_\phi$.  Only models that are isotropic in three
noncompact directions are of interest cosmologically, and so we fix $a=3$.
The Kaluza--Klein reduction of such models
 results in an isotropic universe with $p_1 \dots p_3 = 1/3$, corresponding
to a FRW universe dominated by a component with equation of state $w=1$.

It is important to emphasize that the specific examples that we will discuss are 
only
\emph{representative points} of an open region on the Kasner circle for which 
chaos is 
controlled, chosen so that the Kasner exponents assume a particularly simple,
symmetric form.  
For a given compactification, the selection rules define a reduced set of
stability conditions, which in turn define an open region of the Kasner circle 
for which 
all stability conditions are satisfied.  When this open region is non--empty, 
then 
chaos is controlled.
Thus, there will be choices of the Kasner
exponents with controlled chaos in open neighborhoods of all of the solutions 
discussed
herein.

\subsection{Pure gravity models}

The simplest case, with $n=1$ extra dimensions, is also a somewhat exceptional
one.  This is because there is exactly one compact one dimensional
manifold, the circle $\sphere 1$.  Regardless of the metric on the $\sphere 1$, 
it
will always possess a Killing vector, and so no gravitational stability 
conditions
can be deleted.  Furthermore, $\coH 1 {\sphere 1}$ is
nonzero, and so no $p$--form stability conditions are deleted.  Therefore, all 
chaotic
models remain chaotic when compactified on $\sphere 1$.  To eliminate chaos when 
$n=1$, we must consider a more general class of spaces than manifolds. 

For Einstein gravity without matter,
a simple example that eliminates chaos when $n=1$ is given by the orbifold 
$\smodz$, previously discussed in Ref. \cite{Eri03}. 
If we take a coordinate $\theta$ on $\sphere 1$,
ranging from $[-\pi,\pi]$, then the orbifold results from identifying the 
$\sphere 1$
under the reflection $\theta \to -\theta$.  This takes
 $G_{\mu\theta} \to - G_{\mu \theta}$, and thus the 
Killing field is projected out, giving mass to all the Kaluza--Klein vectors.  
The
resulting action for massless fields in four dimensions is then,
\begin{equation}
S = \int R(g) - (\partial \phi)^2
\, \sqrt{-g} \; d^4 x ,
\end{equation}
in comparison with the classic Kaluza--Klein result \er{eq:5dKK}.
Being only Einstein gravity with a scalar field, this theory is not chaotic.
We will discuss compactification on this particular orbifold in more detail when 
we 
discuss string and M--theory solutions with controlled chaos.

When we have $n>1$ extra dimensions, then chaos can be eliminated by 
compactifying
on a manifold without continuous isometries, and therefore without Killing 
vectors.  
This deletes the mixed 
gravitational stability conditions, as discussed in Section \ref{ss:gravityMSC}.
The remaining stability conditions are always
satisfied in the neighborhood of doubly isotropic solutions.
This is subject to the assumption, discussed at the end of 
\ref{ss:gravitySpectrum},
that the mass matrix for the Kaluza--Klein vector modes has no negative 
eigenvalues.

While we have seen that the masses of Kaluza--Klein vectors are determined by
isometric properties of $\isC$, there is  a  useful class of manifolds
for which these properties are themselves determined by the topology, 
specifically
by the de Rham cohomology.  In this case, the 
gravitational and $p$--form selection rules are determined
entirely by the cohomology of $\isC$.  These are the Einstein manifolds, for
which,
\begin{equation}
R_{MN} = \lambda g_{MN},
\end{equation}
with $\lambda$ arbitrary.  When $\isC$ is Einstein, the number of Killing 
vectors is 
given by $\dim \coH{1}{\isC}$ \cite{Bes87}.  Chaos will thus be controlled in a
neighborhood of doubly isotropic solutions when
$\dim\coH{1}{\isC} = 0$.  There are many examples of Einstein manifolds with 
this 
property; among them are the complex projective spaces $\CP n$ with the 
Fubini--Study metric, and the Calabi--Yau spaces.

\subsection{String models}

For string models with $\mc N = 1$ supersymmetry
in ten dimensions (Type I and heterotic), 
the simple class of doubly isotropic solutions is
sufficient to give examples of solutions with controlled chaos.  
As we will discuss in more detail below,
some of our solutions are related to others through standard string duality
relationships.
Interestingly, we find that theories with $\mc N = 2$ supersymmetry (Type II)
do not admit compactifications that lead to controlled chaos with doubly 
isotropic
solutions.
Unfortunately, we have found no examples where the compactification manifold 
could be a Calabi--Yau, although
solutions with controlled chaos and Calabi--Yau compactification may exist
for non--doubly isotropic choices of the Kasner exponents.

In the following, 
we will always give the Kasner exponents in the Einstein conformal frame. 
Conventionally, the bosonic sector of string theory actions is presented in the
``string frame" form,
\begin{equation}
S_{\text{string}} = \int \left( e^{-2\Phi} \left[ R(G^{(S)}) + 4(\partial 
\Phi)^2 \right] - 
\sum_j e^{ \lambda^{(S)}_j \Phi} F_{p_j+1}^2 \right) \; \sqrt{-G^{(S)}} \, 
d^{10} x
\end{equation}
where the $ \lambda^{(S)}_j$ are the string frame couplings to the dilaton 
field,
and $G^{(S)}$ the string frame metric.
One arrives at the ``Einstein frame" action by the transformation,
\begin{equation}
G_{MN}^{(E)} = e^{-\Phi/2} G_{MN}^{(S)}
\end{equation}
resulting in,
\begin{equation}
S_{\text{Einstein}} = \int \left( R(G^{(E)}) - (\partial \phi)^2 - 
\sum_j e^{ \lambda^{(E)}_j \phi} F_{p_j+1}^2 \right) \; \sqrt{-G^{(E)}} \, 
d^{10} x
\end{equation}
where we have defined,
\begin{subequations}
\begin{align}
\phi &= \Phi/\sqrt{2}, \\
\lambda^{(E)}_j & = \sqrt{2} \left( \lambda^{(S)} + \frac{8 - 2p_j}{4} \right),
\end{align}
\end{subequations}
The  field $\phi$, which we shall refer to as the ``dilaton" below, is 
canonically 
normalized, and the couplings between the
dilaton and $p$--forms have transformed.
In the following, we will always use the Einstein frame couplings, and so will 
drop
the superscript $(E)$ for clarity in notation.

A theme common to our examples is the compatibility of our results 
concerning controlled chaos and the duality relationships connecting various
string theories.
In Section \ref{sss:hetM} we
first examine the $\text{E}_8 \times \text{E}_8$ heterotic theory in detail.  
Through
a combination of string
duality relationships and compactifications we will be able to discuss its 
limits
in eleven, ten, five and four dimensions.   
In particular, the S--duality relating the $\text{E}_8 \times \text{E}_8$ 
heterotic
string and M--theory \cite{Hor95,Hor96} is made apparent by relating the 
ten dimensional heterotic solution with controlled chaos and the 
compactification
of eleven dimensional M--theory on $\sphere{1} / \Zmod{2}$.
We will then discuss all
compactifications of doubly isotropic solutions with controlled chaos for string
theories with $\mc N = 1$ supersymmetry in ten dimensions.  
We give four representative solutions, two each for the heterotic and 
Type I theories.  We show that these four solutions organize into two pairs of
 solutions, related by the
S--duality connecting the heterotic SO(32) and Type I strings \cite{Pol98}.

It is important to keep in mind some features of the space of string solutions 
with
controlled chaos.  Each compactification we discuss,  defined by the 
vanishing de Rham cohomology classes, defines an open region on the Kasner 
circle
where chaos is controlled.  Our restriction to doubly isotropic models, in turn,
takes a one dimensional ``slice" out of this open region.  In our examples, we 
give
a representative point from the ``slice" where the Kasner exponents assume a 
convenient and symmetric form.  Thus we have found the compactifications that 
admit
doubly isotropic solutions, but the choices of Kasner exponents are not unique.

\subsubsection{The heterotic string and M--theory}\label{sss:hetM}

\begin{table}
  \begin{center}
    \begin{tabular}{|| l | l | c | c | c | c | c  ||}
      \hline
      theory & spacetime &
      dim & $p_1 \dots p_3$ & $p_4 \dots p_9$ & $p_{10}$ & $p_\phi$   \\ \hline 
\hline
      M--theory     & $\nonC \times \isC \times \smodz$ &                      
11 & -0.1206  & 0.0662 & 0.9644 &               \\ \hline
      het $\text{E}_8 \times \text{E}_8$          & $\nonC \times \isC$                            
&                      10 & 0        & $1/6$  &        & $\sqrt{5/6}$  \\ \hline
      ``braneworld" & $\nonC \times \smodz$             &                      5  
& 0.0105   &        & 0.9686 & 0.2486        \\ \hline
      FRW           & $\nonC$                                        &                      
4  & 1/3      &        &        & $\sqrt{2/3}$  \\ \hline
      \hline
    \end{tabular}
    \caption{The solution discussed in the text for the $\text{E}_8 \times 
\text{E}_8$ heterotic string in various dimensions.  The Kasner exponent 
$p_\phi$ is a
    combination of the dilaton and volume modulus for the compactified 
dimensions.
    Chaos is controlled provided that $\coH{3}{\isC} = 0$.}
    \label{t:hetMguises}
  \end{center}
\end{table}

Here we focus on the
$\text{E}_8 \times \text{E}_8$ heterotic theory, in the neighborhood of a 
specific
choice of Kasner exponents.  Using string duality relationships and 
compactification,
we will discuss the various 
guises of this solution in eleven, ten, five and four dimensions,
summarized in Table \ref{t:hetMguises}.
While the solution we discuss  also controls chaos for the 
SO(32) heterotic theory in ten dimensions,  string dualities for this theory
do not enable us to discuss the five dimensional and M--theory limits.
To begin, we consider the heterotic theory in Einstein frame, where it contains 
the
metric $G_{MN}$, dilaton $\phi$, one--form $A_1$ and two--form $B_2$.  The 
dilaton
couples to the one-- and two--forms via exponential couplings of the type
\er{eq:pFormDef}, with 
$\lambda_1 = -1/\sqrt{2}$, and $\lambda_2 = -\sqrt{2}$.  
Before compactification, one finds violations of the electric and magnetic 
stability
conditions for the choice of Kasner exponents,
\begin{equation}\label{eq:string_ex_1}
p_1 \dots p_3 = 0, \quad p_4 \dots p_9 = 1/6, \quad p_\phi = \sqrt{5/6} \qquad
\mbox{(10D)}.
\end{equation}
We assume that $p_1 \dots p_3$ lie along the noncompact  spacetime
$\nonC$, and $p_4 \dots p_9$ lie along the compact manifold $\isC$.
In this solution, the magnetic stability conditions are violated for 
the magnetic component of $H_3 = \diff B_2$ with all three indices along $\isC$.
None of the gravitational stability conditions are violated.

Applying the selection rules introduced in Section \ref{s:msc}, we find that
chaos is controlled by compactifying on a six--manifold $\isC$ with
$\coH{3}{\isC} = 0$, such as $\sphere 6$ or
$\CP 3$.  The choice of $\CP 3$ has the advantage that it has
$\coH{1}{\CP 3} = 0$, and is an Einstein manifold if given the Fubini--Study 
metric.  
This manifold may therefore be a useful starting point for models that differ
from the doubly isotropic ones.
Unfortunately, a Calabi--Yau space will always have 
$\dim \coH{3}{\isC} \ne 0$, and is thus unsuitable for rendering this solution 
non
chaotic.

The four dimensional limit of the solution \er{eq:string_ex_1} possesses a 
simple
form.  This is obtained by compactifying on the six--manifold $\isC$, resulting 
in,
\begin{equation}\label{eq:e8e84d}
p_1 \dots p_3 = 1/3, 
\qquad \mbox{(4D)}.
\end{equation}
This describes a collapsing, flat FRW universe dominated by a perfect fluid with
$w=1$.  The $w=1$ component is a combination of the dilaton and
the volume modulus arising from the
Kaluza--Klein reduction of the heterotic theory on the
six--manifold $\isC$.

String duality relationships imply that the 
(strongly coupled) $\text{E}_8 \times \text{E}_8$ heterotic theory is obtained
by compactifying M--theory on the orbifold $\smodz$
\cite{Hor95,Hor96,Luk97}.
Phenomenology implies that the orbifold $\smodz$
is somewhat larger than the compactification
six manifold $\isC$.  Thus, depending on the scale of interest, the strongly
coupled heterotic theory can appear four, five, or eleven dimensional.
The five and eleven dimensional limits of the solution 
\er{eq:string_ex_1} are not as simple as the 
ten and four dimensional views, but are nonetheless instructive.  

This duality relationship implies that the heterotic string in ten dimensions
can be described by  eleven dimensional M--theory on
$\nonC \times \isC \times \smodz$.
The eleven dimensional lifting of \er{eq:string_ex_1} to M--theory yields Kasner
exponents whose precise expression is not very illuminating, but whose
approximate numerical values are,
%
%
\begin{equation}
p_1 \dots p_3 = -0.1206, \quad p_4 \dots p_9 = 0.0662, \quad p_{10} = 0.96442
\qquad \mbox{(11D)}.
\end{equation}
This describes a rapidly shrinking orbifold, a slowly contracting six--manifold 
$\isC$,
and a slowly expanding noncompact space.  

To see that the compactification of M--theory on $\nonC \times \isC \times 
\smodz$
leads to controlled chaos requires us to consider some subtle features of 
the theory.  The bosonic sector of M--theory includes only the graviton and a 
four form field strength $F_4 = \diff A_3$. 
Our selection rules are only strictly applicable
to the case where the compactification space is a manifold, an assumption which 
fails to include orbifolds such as $\smodz$.  The approach most convenient here 
follows
usual techniques \cite{Hor95,Hor96,Luk97} for determining the spectrum after 
compactification.  Specifically, we compactify M--theory on
$\sphere 1$, and then impose the identification $\theta \to -\theta$.  
The presence of the Chern--Simons term $A_3 \wedge F_4 \wedge F_4$ in the M--
theory 
Lagrangian requires that $F_4 \to -F_4$ under parity transformations, of which 
the 
identification $\theta \to -\theta$ is an example.  Thus, the massless 
components
of $F_4$ on $\real{1,3} \times \isC \times \smodz$ 
are those with exactly one index along the $\smodz$.

Before compactification, the eleven--dimensional
 solution given above violates the $p$--form stability
conditions for  three 
components of the four form field.  
The first and second are electric, with the first
having three indices along 
$\nonC$, and the second having two along $\nonC$ and one along $\isC$.
The third is magnetic, with one index along $\smodz$ and three  along 
$\isC$.  The magnetic component is rendered massive by the condition
$\coH{3}{\isC} = 0$, and so we can neglect this stability condition.
The two electric components are rendered massive since they do not have exactly 
one
index along the $\smodz$, and their stability conditions can be neglected as 
well.
Thus, chaos is controlled in the M--theory limit.

The five dimensional guise of our solution, 
obtained by Kaluza--Klein reducing the eleven dimensional form on the six--
manifold
$\isC$, describes a ``braneworld" with structure $\nonC \times \smodz$.
This yields the solution,
\begin{equation}
p_1 \dots p_3 = 0.01048, \quad p_{10} = 0.9686, \quad p_\psi = 0.24804.
\qquad \mbox{(5D)}
\end{equation}
The scalar field $\psi$ is the volume modulus of the six--manifold $\isC$.
This solution describes a nearly static $\nonC$ and a rapidly contracting 
orbifold.  
This solution bears a suggestive 
similarity to the set--up studied in the ekpyrotic/cyclic scenario.
In these models, near the big crunch, the five--dimensional spacetime approaches 
the Milne solution,
\begin{equation}\label{eq:ecKasnerExp}
p_1 \dots p_9 = 0, \quad p_{10} = 1.
\end{equation}
This solution is in fact on the boundary of the open region of the Kasner sphere 
for which our example solution \er{eq:string_ex_1} exhibits controlled chaos.  
This is predicated on the
assumptions that $w=1$ in the four--dimensional theory all the way to the
big crunch, and that compactification is the only mechanism for
controlling chaos.  In the ekpyrotic/cyclic scenarios, there is a long 
$w \gg 1$
phase during the contraction, in which the energy density in $p$--form
modes is exponentially suppressed \cite{Eri03}.  This
suppression further reduces the time $t_{dom}$ at which dangerous modes
can formally dominate the universe.  Therefore, in the full model,
the onset of chaos will
be delayed far beyond what our estimates, based only on the
compactification mechanism, would suggest.

\subsubsection{The heterotic and Type I strings}\label{sss:hetI}

\begin{table}
  \begin{center}
    \begin{tabular}{|| c | c | c | c | c | c ||}
      \hline
      sol'n   &  $p_1 \dots p_3$ & $p_4 \dots p_9$ & $p_\phi$      & theories      
& zero betti \\ \hline \hline
      {\bf A} &  $0$             &  $1/6$          & $\sqrt{5/6}$  & heterotic    
& $b_3$ \\ \hline
      {\bf B} &  $0$             &  $1/6$          & $-\sqrt{5/6}$ & Type I             
& $b_3$ \\ \hline
      {\bf C} &  $1/3$           &  $0$            & $\sqrt{2/3}$  & Type I & 
$b_1,b_2$\\ \hline
       {\bf D} &  $1/3$           &  $0$            & $-\sqrt{2/3}$  & heterotic 
& $b_1,b_2$\\ \hline
     
    \end{tabular}
    \caption{Representative string theory solutions with controlled chaos and 
isotropic behavior along the noncompact and compact directions.  Each 
compactification leads to open regions of the Kasner circle with controlled 
chaos, and we have given a representative point for each open region here.  The 
string theories which exhibit controlled chaos for each solution are shown, as 
well as the Betti numbers 
    $b_j = \dim \coH{j}{\isC}$
    of $\isC$ that are required to vanish.}
    \label{t:stringsolns}
  \end{center}
\end{table}

Having focused in detail on a single compactification of a single string theory,
we now focus on finding all compactifications with controlled chaos and doubly 
isotropic Kasner exponents.  
There are four doubly isotropic examples with controlled chaos,
with representative choices of the Kasner exponents
summarized in Table \ref{t:stringsolns}.
The $\text{E}_8 \times \text{E}_8$ and SO(32) heterotic theories exhibit the 
same
chaotic behavior, since their $p$--form spectrum and couplings to the dilaton 
are
identical; these theories differ only in the gauge groups for their non--abelian
gauge multiplets.  One may also include the 
ten--dimensional (noncritical) bosonic string, which contains only the 
Neveu--Schwarz
fields of the heterotic string and no gauge fields.
One finds that chaos is controlled in the ten dimensional bosonic string in the 
same solutions ({\bf A} and {\bf D}) as in the heterotic string.

In the absence of any compactification, these examples are all chaotic.
None of them suffer from gravitational chaos, and in all cases the chaotic
behavior arises from the $p$--form fields alone.
Upon compactification to four dimensions, these models all result in a 
FRW universe dominated by a free scalar field with $w=1$.

The examples given in Table \ref{t:stringsolns}
include not only models that go to weak coupling at the crunch,
({\bf A} and {\bf C}) but also models where the dilaton runs to strong coupling
({\bf B} and {\bf D}). 
The fact that the solutions include both those where the string theory goes to
strong and weak coupling is interesting from a model building perspective.  
The dilaton is never static in the solutions discussed here, a feature also 
found
in some cosmological models based on string theory.
In the
ekpyrotic/cyclic models, for example, the string coupling goes to zero at the 
big crunch.
In pre big--bang models, on the other hand, the dilaton goes to strong coupling 
at the 
crunch.  Thus the controlled chaos mechanism may be relevant to both scenarios.

The heterotic SO(32) and Type I theories are related by an S--duality 
transformation,
and this symmetry is respected by our examples here.  Under this duality,
the string frame actions of
the heterotic SO(32) and Type I theories are related by \cite{Pol98},
\begin{subequations}
\begin{align}
G_{MN}^{(I)} & = e^{-\Phi_h} G^{(het)} \\
\Phi^{(I)}   & = - \Phi^{(het)}
\end{align}
\end{subequations}
with the $p$--form fields remaining unchanged.  Carefully working through the 
resulting transformation of the Einstein frame Kasner exponents, one finds that 
the
spatial Kasner exponents are unchanged, while $p_\phi^{(I)} = - p_\phi^{(het)}$.
Therefore, $S$--duality exchanges the pairs of solutions 
{\bf A} $\leftrightarrow$ {\bf B} and
{\bf C} $\leftrightarrow$ {\bf D}.

The properties of string theories regarding controlled chaos appear 
correlated to their
supersymmetry properties in ten dimensions.  The $\mc N = 1$ theories, 
(heterotic, Type I, and M--theory on $\smodz$) possess simple 
compactifications that control chaos.  
The Type IIA/B theories and uncompactified M--theory, with $\mc N = 2$ 
supersymmetry, 
have no doubly isotropic solutions with controlled chaos.
As we have not exhaustively examined the Kasner sphere,
we cannot say for certain whether there exist solutions that control chaos for
the $\mc N = 2$ theories.  

It is natural to expect that the $\mc N=1$ and $\mc N=2$ string theories 
will have different characteristics with respect to controlled chaos.  There is 
a
useful formulation of the dynamics of gravity near a big crunch, discussed 
briefly in
Section \ref{s:review}. In this formulation, the dynamics of metric and $p$--
form fields
 is recast as the motion of a billiard ball in a hyperbolic space, undergoing 
reflections from a set of walls.  
The walls correspond to $p$--form kinetic terms and
curvature terms in the Einstein equations.  The positions and orientations of 
these
walls are identical for all of the $\mc N=1$ theories, and different
from the common set of walls shared by the $\mc N=2$ theories \cite{Dam00A}.
Our suppression of
 the energy density in massive $p$--form and gravitational modes amounts
to ``pushing back" these walls.  Thus, it is not surprising that we should find 
that
$\mc N=1$ and $\mc N=2$ models have different characteristics with respect to 
controlling chaos.

\section{Conclusions}\label{s:conclusions}

The results presented here build on the many years of previous research in 
the behavior of general relativity near a  big crunch.  Previous
research has primarily focused on ``local" properties of theories with gravity, 
such as the dimensionality of spacetime, or the 
types and interactions of matter fields,
and has revealed how these influence the emergence of chaos.  Here we
have investigated ``global" features, in particular the topology of spacetime.
We have found that these features can lead to a suppression of chaos in many 
models
of interest.  The control of chaos can be 
expressed simply in terms of selection rules for the gravitational and
$p$--form stability conditions.  These in turn can be used to find 
compactifications of chaotic theories in which chaos is suppressed 
right up to the quantum gravity regime.

Our results bear an intriguing connection to some cosmological models that are
founded on current ideas in string and M--theory.  Among the simple examples
of string theory solutions with controlled chaos, we find those that resemble
both the ekpyrotic/cyclic  and  pre--big bang scenarios.  For future
models, this work suggests a method to control chaotic behavior near a big 
crunch that
does not require postulating additional interactions and matter fields, or 
depending
on higher order corrections to the Einstein equations.  While this work sheds no
light on the behavior of these models in the quantum gravity regime or through 
the
big crunch/big bang transition, it provides a natural mechanism that ensures 
that
the universe evolves smoothly so long as classical physics may be 
trusted.

Recent work suggests that maintaining this smooth contraction during the 
classical
regime may be sufficient to allow a nonsingular quantum evolution through a 
big crunch/big bang transition.  One approach to this problem \cite{TPS04}
begins from the fact
that, in string and M--theory, the degrees of freedom during the quantum 
regime are very different from those of the classical regime studied here.  
The fundamental degrees of freedom are extended objects, such as strings and
branes.  As one approaches the scale set by their tension, classical general
relativity breaks down, and these extended objects become the relevant degrees 
of
freedom.  In particular, it is the evolution of these strings and branes that 
one 
should study near the big crunch.  
Working within the context of the ekpyrotic/cyclic scenario, it was
found in Ref \cite{TPS04}
that if the universe is sufficiently smooth and homogeneous at the beginning of
the quantum regime, the fundamental excitations (M2 branes) evolve smoothly 
through the
big crunch with negligible backreaction.  This suggests that a sufficiently 
smooth
``in" state can evolve through the big crunch to a smooth ``out" state, 
precisely what
one requires for cosmology.  This result complements the present work.  
The mechanism described herein can be viewed as providing the required 
conditions 
for  smooth
classical evolution before the Einstein equations break down, preparing the 
universe for nonsingular quantum evolution through the big crunch.

Our results have further implications for high energy theory and phenomenology.  
String models and M--theory require compactification in order to produce the 
correct 
number of observed noncompact dimensions.  Obtaining
the correct low energy physics, such as $\mc N=1$ supersymmetry in four 
dimensions
or the correct number of lepton generations, puts constraints on the
compactification manifold $\isC$, many of which are topological in nature.  
Controlling chaos through compactification in cosmological models with a 
collapsing phase places additional constraints
on $\isC$.  We are currently investigating whether these two set of constraints 
are 
compatible.  For example, the existence of 
solutions with compactification on a Calabi--Yau
space would suggest that chaos can be controlled in string models with a 
realistic low
energy spectrum.

These results also inspire more speculative scenarios.  
When the universe enters a chaotic regime,
the Kasner exponents will undergo an infinite number of  ``jumps" to different 
points 
on the Kasner sphere as the big crunch is approached.
We also might expect that the topology of $\isC$ is changing at the same time.
For example, there are
situations in string theory where the topology of $\isC$ can change dynamically, 
such as the conifold or flop transitions.      
If the combination of Kasner exponents and topology
lead to controlled chaos, then the universe will subsequently contract smoothly 
to
the big crunch.  In this way, the universe will have dynamically selected not 
only
some properties of $\isC$, but also a ``preferred" cosmological solution near 
the
big crunch.  Analysis of such a scenario would require a much deeper 
understanding of 
cosmology in the quantum gravity regime than is currently
available, clearly an important topic for further research.

\begin{acknowledgments}
DHW would like to thank Daniel Baumann for a careful reading of this manuscript.
This work was supported in part by an NSF Graduate Research Fellowship (DHW),
by  US Department of Energy Grant DE-FG02-91ER40671 (PJS), and by PPARC (NT).
\end{acknowledgments}

\section*{Appendix 1: Hamiltonian Formulation of $p$--form Dynamics}

Below, we treat the case of a general $p$--form field with 
coupling to the dilaton $\phi$, in a collapsing universe.  
We concern ourselves with the case where spacetime is isotropic, as
this is the cosmologically relevant situation after compactification. 
We show that far from the big crunch, 
the energy density in massive fields evolves like that of a pressureless fluid,
$\rho \sim 1/\mbox{(comoving volume)}$.  We will first recast the $p$--form 
dynamics in Hamiltonian
form.  This  allows us to apply the virial theorem and stress energy
conservation to obtain the scaling in energy density far from the crunch.  
The $p$--form action with mass term and dilaton coupling is,
\begin{equation}
S = - \frac{1}{(p+1)!} \int 
\left( \diff A \cdot \diff A 
+ m^2  A \cdot A  \right) e^{\lambda\phi} \, \sqrt{-G} \; d^{D}x,
\end{equation}
where we have fixed the coordinate gauge so that
$ds^2 = -n^2 dt^2 + \gamma_{jk} dx^j dx^k$.
We  choose the canonical coordinates to be the gauge potential 
$A_{\alpha_1 \dots \alpha_p}$.  
The corresponding canonical momenta are,
\begin{equation}
\Pi^{j_1 \dots j_p} = - F^{t j_1 \dots j_p} n e^{\lambda\phi}
\sqrt{\gamma}
\end{equation}
Passing to the Hamiltonian, we find,
\begin{equation}
H = \frac{1}{2} \int \tilde n \left( \Pi \cdot \Pi
+ \gamma e^{2\lambda\phi} F^{(B)} \cdot F^{(B)} 
+ \gamma e^{2\lambda\phi} m^2 A \cdot A
- \gamma e^{2\lambda\phi} A_{t\alpha_2 \dots \alpha_p} \partial_a \Pi^{a 
\alpha_2 \dots \alpha_p} \right) \; d^{D-1} x
\end{equation}
where we use a rescaled lapse function $\tilde n = n e^{-\lambda \phi}/ 
\sqrt{\gamma}$,
and denote the magnetic components of $F_p$ by $F^{(B)}$.  Dot products are 
taken
with respect to the metric $\gamma_{MN}$.
The last term in the integral shows that the ``electric" gauge field modes 
$A_{tj_2 \dots j_{p-1}}$ appear as Lagrange multipliers necessary to enforce
the Gauss's law constraint, but are otherwise 
nondynamical \cite{Dir64}.  We will choose the 
Coloumb gauge, in which $A_{t j_2 \dots j_p} = 0,$ and drop this constraint
term from now on.

The Hamiltonian is in fact exactly that of a set of simple harmonic oscillators.
After decomposing the functions $A_p$ in an appropriate orthonormal set of 
Fourier components, different Fourier modes decouple and the Hamiltonian is
quadratic in $\Pi$ and $A$.
The electric field modes appear
as the kinetic terms, and the magnetic field and mass 
terms correspond to the potential of the
oscillators.
The oscillator potential is 
time dependent, both due to the appearance of 
$e^{2\lambda\phi}\gamma$ and individual metric components in the
magnetic and mass terms.

We are primarily interested in the dynamics of the $p$--form far from the 
big crunch.  In this regime, we
may view the changing scale factors as slowly varying parameters in our 
Hamiltonian.
They will change the spring constants on a timescale given by $t$, the proper 
time
to the big crunch.  
The dynamical timescale (typical period) for the oscillator
 Hamiltonian is given by the mass 
term.  Thus, we expect that the fractional change in the fundamental frequencies
$\omega$ of the oscillator system over a 
typical cycle will be, 
\begin{equation}\label{eq:adiabatic}
\frac{ \delta \omega }{\omega} \sim \frac{1}{\omega t}
\end{equation}
Provided we are at a time $t \gg \omega^{-1}$, we will be in the adiabatic 
regime, and
we may take the oscillation frequencies to be constants.  This corresponds
to the $m/H \gg 1$ regime that we have concerned ourselves with throughout this
paper.

To find how the energy density scales with time, we apply the virial theorem and
stress energy conservation.  The adiabatic condition \er{eq:adiabatic} implies
that we may neglect the time variation of the metric and dilaton over a single
cycle.  For
our Hamiltonian, the virial theorem then implies that the time average 
$\langle \cdot \rangle$ of the potential energy is equal
to that of the kinetic energy, or
\begin{equation}\label{eq:pFormVirial}
\langle \Pi \cdot \Pi \rangle = e^{2\lambda\phi}\gamma \langle
F^{(B)} \cdot F^{(B)} 
+ m^2 A \cdot A \rangle .
\end{equation}
In the virialized system, there are two possible regimes, corresponding to 
either
the $F^{(B)}\cdot F^{(B)}$ term or the $m^2 A \cdot A$ term dominating.
We will consider both of these cases in turn.

The stress energy for the $p$--form field is,
\begin{equation}
{T_\mu}^\nu = \frac{e^{\lambda \phi}}{(p+1)!} \left[
(p+1) F_{\mu \alpha_2 \dots \alpha_{p}}F^{\nu \alpha_2 \dots \alpha_{p}}
- \frac{1}{2} {\delta_\mu}^\nu F^2 + 
p m^2 A_{\mu \alpha_2 \dots \alpha_{p-1}}A^{\nu \alpha_2 \dots \alpha_{p-1}}
- \frac{m^2}{2} {\delta_\mu}^\nu A^2
\right]
\end{equation}
We will find it convenient to break the stress energy into three parts, 
\begin{equation}
{T_\mu}^\nu = {{T^{(E)}}_\mu}^\nu + {{T^{(B)}}_\mu}^\nu + {{T^{(m)}}_\mu}^\nu
\end{equation}
corresponding
to the energy in electric modes, magnetic modes, and the mass term.
It is sufficient to consider a single component of $F_p$, since different 
components
will be uncorrelated and therefore will have vanishing time average.  
The electric modes give rise to a contribution,
\begin{equation}
{{T^{(E)}}_\mu}^\nu = {\delta_\mu}^\nu
\frac{e^{\lambda \phi}}{2(p+1)!} | F_{(E)}^2 | \times
\begin{cases}
-1 & \qquad \text{if } F^{(E)} \text{ has index } \mu, \\
+1 & \qquad \text{otherwise}.
\end{cases}
\end{equation}
while the magnetic modes give,
\begin{equation}
{{T^{(B)}}_\mu}^\nu = {\delta_\mu}^\nu
\frac{e^{\lambda \phi}}{2(p+1)!} F_{(B)}^2 \times
\begin{cases}
+1 & \qquad \text{if } F^{(B)} \text{ has index } \mu, \\
-1 & \qquad \text{otherwise}.
\end{cases}
\end{equation}
and the mass term yields,
\begin{equation}
{{T^{(m)}}_\mu}^\nu = {\delta_\mu}^\nu
\frac{e^{\lambda \phi}}{2(p+1)!} m^2 A^2 \times
\begin{cases}
+1 & \qquad \text{if } A \text{ has index } \mu, \\
-1 & \qquad \text{otherwise}.
\end{cases}
\end{equation}
Note that $F^{(B)}$ cannot have any timelike indices, nor can $A$ thanks to our
gauge choice.  Thus the contributions to the energy density
$\rho = -{T_0}^0 $ are all positive.

First, we consider the case where the mass term dominates in the
virial relationship \er{eq:pFormVirial}.  This corresponds to 
inhomogeneities in the $p$--form field being negligible. 
 The virial result implies that 
 $\langle | F_{(E)}^2 | \rangle = m^2 \langle  A^2 \rangle$.
 Due to our gauge choice,
$A$ and $F^{(E)}$ have the same combination of $p$ spatial indices, and 
therefore
contributions to
the pressure components ${T_j}^j$ coming from ${{T^{(E)}}_j}^j$
and ${{T^{(m)}}_j}^j$ exactly cancel.
The vanishing pressure reveals that the effective equation of state is that of
dust, $w=0$.

When the magentic terms dominate the virial result \er{eq:pFormVirial}, we 
obtain
a slightly different effective equation of state.  Here it is necessary to 
average
over polarizations of $F_p$, since 
unlike $F^{(E)}$ and $A$, $F^{(E)}$ and $F^{(B)}$ do not enjoy any
relationships between their indices.  Regardless of polarization, the 
sum of stress energy tensors $T^{(E)}$ and $T^{(B)}$ has vanishing trace.  This,
combined with isotropy, implies the pressure components are given by
${T_j}^j = - {T_0}^0 / (D-1)$.  This corresponds to the equation of state of
radiation, which in four dimensional spacetime is $w=1/3$.

Physically, this result may be understood in simple terms.  Far from the big 
crunch,
the contraction of space is very slow in comparison to the mass of the $p$--form 
field.
Thus, the corresponding particles are far from being relativistic, and behave as 
a dust of approximately comoving mass points.  Their energy density therefore 
scales
in inverse proportion to the comoving volume.  The case where magnetic 
components
dominate the virial relationship corresponds to a relativistic gas of particles.
This yields the equation of state of radiation, as we expect.

\bibliographystyle{unsrt}

\begin{thebibliography}{99}
\bibitem{conventions}
Our conventions are $8\pi G=c=1$, and 
$(-+\cdots+)$ sign convention for the metric.
\end{thebibliography}

\end{document}